\documentclass[12pt]{article}
\usepackage{amssymb}
\usepackage{amsfonts}
\usepackage{amsmath}
\usepackage{graphicx}
\usepackage{color}

\def\Z{\mathbb Z}

\def\R{\mathbb R}

\newcommand{\beq}{\begin{equation}}
\newcommand{\eeq}{\end{equation}}
\newcommand{\bea}{\begin{eqnarray}}
\newcommand{\eea}{\end{eqnarray}}

     \advance \textheight by \topskip
     \makeatletter
     \@addtoreset{equation}{section}

     \makeatother
\def\beq{\begin{equation}}
\def\eq{\end{equation}}

\def\12{\frac{1}{2}}

\begin{document}

\title{ {\bf Self-isospectral tri-supersymmetry
in ${\cal PT}$-symmetric quantum systems  with pure imaginary
periodicity }}

\author{
{\bf Francisco Correa${}^{a}$ and Mikhail S. Plyushchay${}^{b}$}
\\
[4pt]
{\small \textit{${}^{a}$ Centro de Estudios Cient\'{\i}ficos (CECs), Casilla 1469, Valdivia, Chile}}\\
{\small \textit{${}^{b}$ Departamento de F\'{\i}sica, Universidad de
Santiago de Chile, Casilla 307, Santiago 2,
Chile  }}\\
\sl{\small{E-mails: correa@cecs.cl, mikhail.plyushchay@usach.cl} }}
\date{}

\maketitle

\begin{abstract}
We study a  reflectionless $\mathcal{PT}$-symmetric quantum system
described by the pair of complexified Scarf II potentials mutually
displaced in the half of their pure imaginary period. Analyzing the
rich set of intertwining discrete symmetries of the pair, we find an
exotic supersymmetric structure based on three matrix differential
operators that encode all the properties of the system, including
its reflectionless (finite-gap) nature. The structure we revealed
particularly sheds new light on the splitting of the discrete states
into two families, related to the bound and resonance states in
Hermitian Scarf II counterpart systems, on which two different
series of irreducible representations of $sl(2,\mathbb{C})$ are
realized.
\end{abstract}


\section{Introduction}

Fourteen years ago,  Bender and Boettcher discovered a huge and
remarkable class of non-Hermitian quantum Hamiltonians which exhibit
an entirely real spectrum \cite{B&B}. One of the key points of the
observation was the requirement of a $\mathcal{PT}$-\emph{symmetry}
generated by the product of the parity, ${\cal P}$, and time, ${\cal
T}$, inversion operators,
\begin{equation}\label{PTdef}
    {\cal P}\,x\,{\cal P}=-x,\qquad
    {\cal T}\,x\,{\cal T}=x,\quad
    {\cal T}\,i\,{\cal T}=-i\,,
\end{equation}
 that substitutes the usual quantum mechanical property of
Hermiticity. This condition, however, is necessary but not
sufficient for the reality of the spectrum; it is also required that
eigenfunctions of a ${\cal PT}$-symmetric Hamiltonian must be
simultaneously eigenfunctions of the $\mathcal{PT}$ operator. In
this case we say that the ${\cal PT}$ symmetry is unbroken,
otherwise the eigenvalues are not real, a part or all of them appear
in complex conjugate pairs, and the ${\cal PT}$ symmetry is broken
\cite{B&B,BBM,DDT}.

Nowadays, quantum mechanics with non-Hermitian Hamiltonians
transformed into an independent line of research where,
specifically, the notion of the ${\cal PT}$-symmetry was generalized
into the condition of pseudo-Hermiticity \cite{Mpseudoh}.
Non-Hermitian Hamiltonians appear in physics in diverse areas
including   quantum optics, cosmology, atomic and condensed matter
physics, magnetohydrodynamics, among others. For a good review of
the developments in the area and applications, see
\cite{BenderReview, ReviewM} and references therein.

On the other hand, there is a wide class of modern techniques and
methods which prove their effectiveness in the study of quantum
mechanical systems. One of them is supersymmetric quantum mechanics
(SUSYQM), introduced initially by Witten as a toy model to study the
spontaneous supersymmetry breaking \cite{witten}. Over the past few
decades, SUSYQM  transformed into a powerful tool in quantum physics
\cite{susyrev1} which turns out to be useful, for example, in  the
spectral analysis as well as in searching for  new solvable systems.
SUSYQM in its usual form is based on a Darboux transformation that
relates two Hamiltonians by means of an intertwining, linear
differential operator, and leads, as a result, to a complete or
almost complete isospectrality of both systems \cite{MatSal}.

The concept of Darboux transformations and SUSYQM was generalized in
different aspects that lead to the discovery of the classes of
quantum systems that reveal  non-linear \cite{nSUSY, MPhidnon,KlMP},
bosonized \cite{MPhidnon, bosonized,  boso1, boso2} and
self-isospectral \cite{Dunne, Fetal, FerNegNie, self, trisusy}
supersymmetries. Particularly, a certain class of potentials was
found in which all the mentioned specific types of supersymmetries
were brought to light in a form of a peculiar structure that was
coined in \cite{trisusy} as ``\emph{tri-supersymmetry}". Such a
structure was shown to underlie special properties of some physical
quantum systems \cite{trisusydelta, ABtrisusy,
newself1,newself2,newself3}. Among the principal characteristics of
tri-supersymmetry, which will be described below, is the existence
of  three integrals of motion in the form of supercharges which
encode the main properties of the corresponding systems. An example
of the systems that reveal a tri-supersymmetric structure is
provided by the Hermitian \emph{finite-gap} periodic potentials
\cite{trisusy}, which in the limit when their real period tends to
infinity are known as \emph{reflectionless} potentials.

The idea of supersymmetry in some of its versions was adapted in the
context of non-Hermitian Hamiltonians \cite{susypt1, susypt2,
susypt3, susypt4, Znojil2000, susypt5, Dorey, Mpseudosusy}. Although
the non-linear and self-isospectral supersymmetries were studied
before in the systems with non-Hermitian Hamiltonians
\cite{susynonlinearpt}, the presence in them of a supersymmetric
structure that would unify the mentioned types of supersymmetry
remains to be unknown. One can wonder therefore if there exist
non-Hermitian potentials of finite-gap nature that display the
properties to be similar to those of the Hermitian counterpart, and
what are the peculiarities of the associated supersymmetric
structure.

The purpose of the present article is to report the observation of
an exotic tri-supersymmetric structure in a broad class of ${\cal
PT}$-symmetric extended finite-gap systems. This allows us, on the
one hand, to understand and explain the main features of their
spectra by analyzing the properties of the supercharges; on the
other hand, we explicitly trace out the differences that appear in
comparison with the Hermitian case, at the level of potentials as
well as in the generic properties of tri-supersymmetry. The
importance of a hidden, pure imaginary period in non-periodic on a
real line finite-gap systems is clarified, particularly, in the
light of tri-supersymmetric structure.   We achieve all this  by
considering the pairs of ${\cal PT}$-symmetric complexified Scarf II
potentials mutually displaced in the half of their unique imaginary
period. For special values of the parameters these potentials become
perfectly transparent, i.e. reflectionless. It is exactly for such
special parameter values the tri-supersymmetric structure arises and
allows us to clarify its nontrivial interplay with various discrete
symmetries of the potentials. The supersymmetric structure we reveal
gives us a new insight on the splitting of discrete states in such a
class of $\mathcal{PT}$-symmetric quantum systems into two distinct
families, on which two different representations of the
$sl(2,\mathbb{C})$ algebra are realized, the fact that was
established initially by Bagchi and Quesne by using group
theoretical methods \cite{GrNe}. Note here that the complexified
Scarf II potential was extensively studied in the literature in  the
context of ${\cal PT}$-symmetric and pseudo-Hermitian Hamiltonians;
the updated summary on these investigations can be found in ref.
\cite{BagQues}. Recently, this potential also attracted the
attention in different areas of physics such as quantum field theory
in curved spacetimes \cite{Maloney}, soliton
theory in nonlinear integrable systems \cite{Wadati} and also the
physics of optical solitons \cite{Optics}.

The plan of the article is as follows. In sections \ref{sec1} and
\ref{eigen} the basic properties of the pair of complexified
mutually conjugated Scarf II potentials are reviewed. Specifically,
in the next section we study various discrete symmetries of the
potentials, describe spectral properties of the systems in
dependence on the parameter values, and describe the relation with
the other known potentials. In section \ref{eigen} we first present
the explicit expressions for the two families of the singlet states
in the spectrum, and then discuss the continuous spectrum and its
relation with that of the free particle by means of Darboux-Crum
transformations that underlie the non-linear supersymmetry. The
tri-supersymmetric structure is described in section \ref{susysec}.
Section \ref{examples} provides two concrete nontrivial examples of
the systems with two and three bound states to illustrate the
general results. In section \ref{discus} we present the discussion
and concluding remarks.

\section{Discrete symmetries and relations}\label{sec1}

\begin{figure}[h!]
\centering
\includegraphics[scale=0.9]{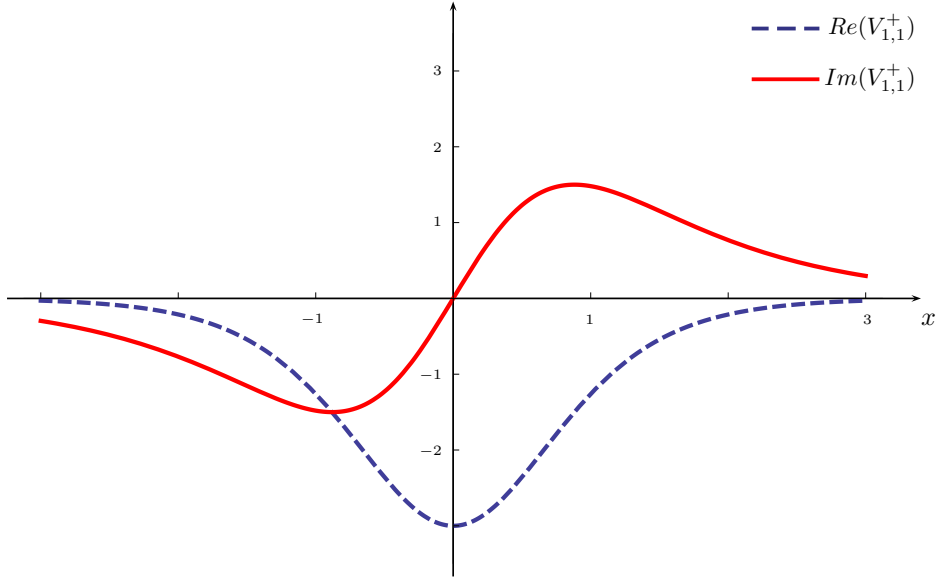}
\caption{Plot of real and imaginary parts of the complexified Scarf
II potential $V^+_{1,1}$, which here as well as in a generic case of
(\ref{po1})  are even and odd functions, respectively.}
\label{FigaSc}
\end{figure}

\begin{figure}[h!]
\centering
\includegraphics[scale=0.9]{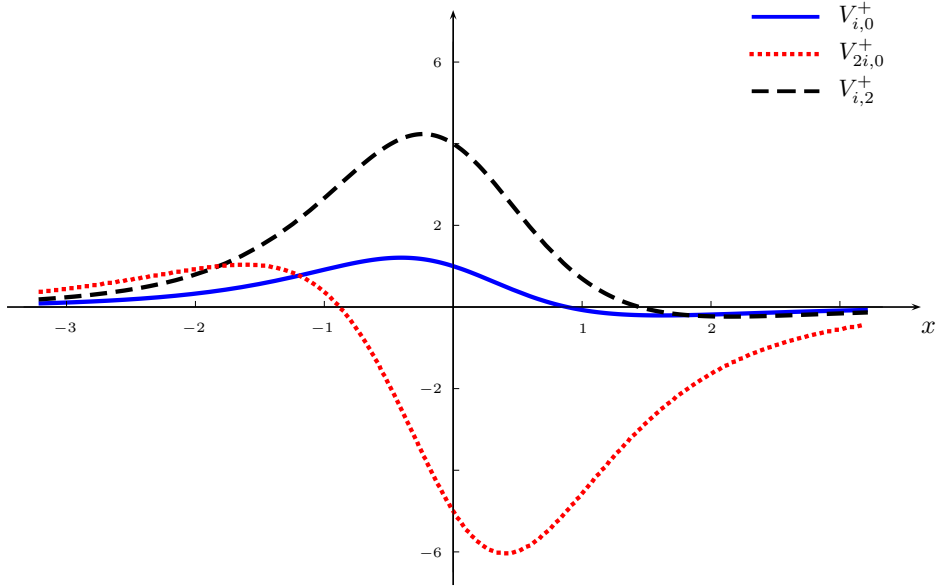}
\caption{Examples of original (real) Scarf II potentials.}
\label{Scarf}
\end{figure}

Consider a pair of  complexified Scarf II potentials
\begin{equation}\label{po1}
    V_{l,m}^{\pm}(x)=-\frac{l^2+m(m+1)}{\cosh^2x}\pm i l
    (2m+1)\frac{\sinh x}
    {\cosh^2 x} \,\, .
\end{equation}
We suppose  that  $l$ and $m$ are real parameters, and $x\in\R$.
Potentials (\ref{po1}) are then free of singularities on the real
line, and their real and imaginary parts vanish for $x\rightarrow
\pm \infty$, see Fig.~\ref{FigaSc}.
Pair (\ref{po1}) can be transformed into a pair of original, real
Scarf II potentials \cite{Scarf2} by a substitution $l\rightarrow
il$, see Fig. \ref{Scarf}.
Both the original and complexified  potentials have a pure imaginary
period $2i\pi$.

Parity inversion, ${\cal P}$, and
time reversal, ${\cal T}$, operators intertwine the completely
isospectral potentials (\ref{po1}),
\begin{equation}\label{pt1}
    {\cal P}V_{l,m}^{\pm}=V_{l,m}^{\mp}{\cal P},
    \qquad {\cal
    T}V_{l,m}^{\pm}=V_{l,m}^{\mp}{\cal
    T} .
\end{equation}
As $V^\pm_{-l,m}=V^\pm_{l,-m-1}=V^\mp_{l,m}$, another pair of
intertwiners is provided by operators $\mathcal{R}_l$ and
$\mathcal{R}_m$,  which act on the parameters $l$ and $m$,
$\mathcal{R}_l:\,(l,m)\rightarrow (-l,m)$,
$\mathcal{R}_m : (l,m) \rightarrow (l,-m-1)$,
    \begin{equation}\label{Rlm}
    \mathcal{R}_l V_{l,m}^{\pm}=V_{l,m}^{\mp}
    \mathcal{R}_l,
    \qquad
    \mathcal{R}_m V_{l,m}^{\pm}=
    V_{l,m}^{\mp}\mathcal{R}_m\,.
\end{equation}
Yet another pair of intertwining operators is given by pure
imaginary translations for the half of the  period $2i\pi$, $T_+:\
x\rightarrow x+i\pi$ and $T_-:\ x\rightarrow x-i\pi$,
\begin{equation}\label{T+-}
    T_+ V_{l,m}^{\pm}=V_{l,m}^{\mp}
    T_+,
    \qquad
    T_- V_{l,m}^{\pm}=
    V_{l,m}^{\mp}T_-\,.
\end{equation}
The product of any two of the listed intertwining operators, except
of $T_+=\exp(\pi\frac{d}{dx})$ and $T_-=\exp(-\pi\frac{d}{dx})$,
$T_+T_-=1$,  is a nontrivial discrete symmetry of each of potentials
(\ref{po1}). Particularly, we find that each potential of the pair
is  ${\cal PT}$-symmetric,
\begin{equation}\label{PTsym}
[{\cal PT},V_{l,m}^{\pm}]=0\,.
\end{equation}
The potentials satisfy also the relation
\begin{equation}\label{dis1}
    V_{l,m}^\pm=V_{-l,-m-1}^\pm
    \,,
\end{equation}
which is produced by a composition $\mathcal{R}_l\mathcal{R}_m$ of
the intertwining generators (\ref{Rlm}).

We also have a symmetry relation
\begin{equation}\label{dis1+}
    V_{l,m}^\pm=V_{m+ \frac{1}{2},\,
    l-\frac{1}{2}}^\pm\,,
\end{equation}
that will play a key role in the analysis below. The composition of
(\ref{dis1}) and (\ref{dis1+}) produces yet another symmetry
\begin{equation}\label{dis2}
    V_{l,m}^\pm=V_{-m-\frac{1}{2},
    -l-\frac{1}{2}}^\pm \,\, .
\end{equation}

The nature of relations (\ref{Rlm}) and (\ref{dis1})--(\ref{dis2})
takes a somewhat more transparent form if to redefine the parameters
\cite{LeZno}: $(l,m)\rightarrow (\alpha_+,\alpha_-$),
$\alpha_\pm=m\pm l+\frac{1}{2}$. Then the coefficients in
(\ref{po1}) are transformed into $l^2+m(m+1)=\frac{1}{2} [\alpha_+^2
+\alpha_-^2-\frac{1}{2}]$ and $l(2m+1)=\frac{1}{2}(\alpha_+^2
-\alpha_-^2)$, and symmetry (\ref{dis1+}) corresponds to a
reflection $\mathcal{P}_{\alpha_-}:\,(\alpha_+,\alpha_-)\rightarrow
(\alpha_+,-\alpha_-)$ in the plane of $\alpha$-parameters.
Intertwining relations (\ref{Rlm}) are given by the products of
reflection $\mathcal{P}_{\alpha_-}$ and $\alpha$-rotations for
$\pm\pi/2$, $\mathcal{R}_{+\pi/2}:\,(\alpha_+,\alpha_-)\rightarrow
(\alpha_-,-\alpha_+)$ and
$\mathcal{R}_{-\pi/2}:\,(\alpha_+,\alpha_-)\rightarrow
(-\alpha_-,\alpha_+)$,
$\mathcal{R}_l=\mathcal{P}_{\alpha_-}\mathcal{R}_{+\pi/2}$,
$\mathcal{R}_m=\mathcal{P}_{\alpha_-}\mathcal{R}_{-\pi/2}$, while
symmetries (\ref{dis1}) and (\ref{dis2}) correspond, respectively,
to a $\pi$-rotation $(\alpha_+,\alpha_-)\rightarrow
(-\alpha_+,-\alpha_-)$ and to a reflection in $\alpha_+$,
$\mathcal{P}_{\alpha_+}:\,(\alpha_+,\alpha_-)\rightarrow
(-\alpha_+,\alpha_-)$.

It will be more convenient for us, however, to work in terms of the
parameters $l$ and $m$. Relations (\ref{dis1})--(\ref{dis2}) allow
us to interchange integer values with half-integer ones as well as
positive with negative values. By this reason, without loss of
generality we can suppose that $l$ and $m$ are non-negative
integers; sometimes, however, negative and half-integer values will
be important as well.

The described discrete relations and symmetries form the base for a
rich exotic supersymmetric structure that we will reveal in the
extended system ${\cal H}_{l,m}=diag(H^{+}_{l,m},H^{-}_{l,m})$,
\begin{equation}\label{hamil}
    H^{\pm}_{l,m}=-\frac{d^2}{dx^2}+V_{l,m}^{\pm}\,,
\end{equation}
with pseudo-Hermitian Hamiltonians \cite{Mpseudoh}\footnote{A pair
of complex potentials (\ref{po1}) with $m=0$ and corresponding
extended Hamiltonian ${\cal H}_{l,0}$  emerged recently in
\cite{Maloney} under investigation of quantum field theory in de
Sitter space.},
\begin{equation}
    H^{\pm}_{l,m}={\cal P}
    (H^{\pm}_{l,m})^\dagger\,{\cal P}\, ,
\end{equation}
where we have taken into account the
relation ${\cal P}^{-1}={\cal
P}$.

The spectra of $H^{\pm}_{l,m}$ display different features in
dependence on the values of the parameters $l$ and $m$. The
eigenfunctions of Hamiltonians (\ref{hamil}) may  or may not be
simultaneous eigenstates of the ${\cal PT}$ operator. The situation
is known as unbroken ${\cal PT}$ symmetry in the former case, or
spontaneously broken in the latter case. As was shown in \cite{B&B,
BBM, DDT}, the spectrum of a ${\cal PT}$ symmetric Hamiltonian is
real when its eigenfunctions are simultaneously the eigenstates of
the ${\cal PT}$ operator. In the case of the $H^{\pm}_{l,m}$, this
situation holds when
\begin{equation}
    l^2+\left(m+1/2\right)^2\geq |2l\left(m+1/2\right)|\,,
\end{equation}
that is always valid for real $l$ and $m$  \cite{Ahmed}. For real
$l$ and $m$, the spectra of $H^{+}_{l,m}$  and $H^{-}_{l,m}$ have a
finite number of bound states of non-degenerate energies, while the
probability flux is conserved in a scattering sector. It was pointed
out in \cite{LevaiCanVen2} that an imaginary displacement in the
spatial coordinate $x \rightarrow x+i\delta$, where $\delta \in
\mathbb{R}$, breaks in general the conservation of probability, i.e.
$|T|^2+|R|^2\neq1$. There is, however,  a case that shows several
special properties for this class of shifted potentials. When $l$
and $m$ take simultaneously non-negative integer values, the
reflection coefficient vanishes, $|R|=0$, and the probability is
conserved independently of any displacement $\delta\neq \pm \pi/2$,
see below. As a result potentials (\ref{po1}) belong to the class of
$ \emph{reflectionless}$ potentials \cite{reflectionless}. In
this case both potentials have the same spectrum with $m +l+1$
singlet states, where $m+l$ of them are bound states of negative
energies, and there is a singlet state of zero energy at the bottom
of the continuous part of the spectrum. We discuss the spectral
characteristics of (\ref{hamil}) in more details in Section
\ref{eigen}.

Within a framework of $\mathcal{PT}$-symmetric quantum mechanics
\cite{BenderReview}, or in a more general framework of the systems with
pseudo-Hermitian Hamiltonians \cite{ReviewM}, it was shown that the
$\mathcal{PT}$ integral can be supplied with yet another nontrivial
(nonlocal) integral of motion that has a nature of the charge
conjugation operator, $\mathcal{C}$ \cite{Coperator}. This allows finally to define a
positive definite scalar product to extend  a probabilistic
interpretation for the case of $\mathcal{PT}$-symmetric systems.
Here, we just have in mind this general picture when discuss the
bound and scattering states by referring to original papers on the
subject.

The double degeneracy in the spectrum for the scattering sector
together with the reflectionless property and the finite number of
singlet states indicate that each system $H^{+}_{l,m}$ and
$H^{-}_{l,m}$ possesses a hidden, bosonized non-linear supersymmetry
\cite{MPhidnon, bosonized} as this happens for Hermitian
reflectionless Hamiltonians \cite{boso1, trisusy}. In fact for
integer values of $l$ and $m$, potentials (\ref{po1}) are solutions
of the KdV hierarchy: they satisfy the stationary $s$-KdV$_{n}$,
$n=l+m$, non-linear equations \cite{KdV}. This means that there
should exist non-trivial integrals of motion  ${\mathbb
A}^{+}_{2n+1}$ and ${\mathbb A}^{-}_{2n+1}$ for  Hamiltonians
(\ref{hamil})  in the form of differential operators of order $2n
+1$. These integrals should satisfy relations
\begin{equation}\label{Adef}
    [{\mathbb A}^{+}_{2n+1},H^{+}_{l,m}]=[{\mathbb A}^{-}_{2n
    +1},H^{-}_{l,m}]=0,
    \qquad ({\mathbb A}^{\pm}_{2n
    +1})^2=P(H^{\pm}_{l,m}),\qquad n=l+m\,,
\end{equation}
where $P(H^{\pm}_{l,m})$ is a polynomial of order $2n+1$, i.e.
${\mathbb A}^{+}_{2n+1}$ and $H^{+}_{l,m}$, as well as ${\mathbb
A}^{-}_{2n+1}$ and $H^{-}_{l,m}$, should compose a Lax pair
\cite{Lax,Belo}.

There are particular cases for which $H^{\pm}_{l,m}$ are completely
isospectral to the reflectionless P\"oschl-Teller Hamiltonians
\cite{PT}\footnote{The free particle  corresponds to $\lambda=0$ and
$\lambda=-1$, and can be considered as a zero-gap case of the family
of finite-gap, reflectionless P\"oschl-Teller systems
(\ref{PT}) \cite{boso1, ptads}.}
\begin{equation}\label{PT}
    H^{PT}_{\lambda}=-\frac{d^2}{dx^2}-
    \frac{\lambda(\lambda+1)}{\cosh^2 x},
    \qquad \lambda \in \mathbb{Z}\, .
\end{equation}
Namely, the P\"oschl-Teller and the complexified Scarf II potentials
are related by  a transformation of an imaginary displacement  and a
rescaling of the variable,
\begin{equation}\label{ptrela}
    H^{PT}_{\lambda}(x\pm i\pi/4)=4H^{\pm}_{l,m}(2x)\,,\quad \text{for}
    \quad
    \left\{
    \begin{array}{ccc}
      m=l-1, \quad   \quad \lambda=m+1\,,    \\
      \text{and}\\
    m=l, \quad    \quad \lambda=m\,.
    \end{array}
     \right.
\end{equation}
Another interesting relation is with a generalized, singular
P\"oschl-Teller potential \cite{susyrev1}
\begin{equation}\label{gpt}
    V_{l,m}^{GPT}(x)=\frac{l^2+m(m +1)}{\sinh^2 x}+l(2m+1)\frac{\cosh x}
    {\sinh^2 x} \, .
\end{equation}
The singularity can be removed by complex shifting \cite{GrNe}. For
special values of such a shifting we get the ${\cal PT}$-symmetric
pair of potentials (\ref{po1}),
\begin{equation}
    V_{l,m}^{GPT}(x\pm i\pi/2)=V_{l,m}^{\pm}(x)\, .
\end{equation}

It is worth to note that the potentials (\ref{po1}) and
corresponding Hamiltonians admit a representation that generalizes
Eq. (\ref{ptrela}),
\begin{equation}\label{h1}
    H_{l,m}^+=\frac{1}{4} \left(-
    \frac{d^2}{d\xi^2}-\frac{r(r+1)}{\cosh^2 \xi}- \frac{s(s+1)}{\cosh^2
    (\xi+i\frac{\pi}{2})}  \right) \,,
\end{equation}
\begin{equation}\label{h2}
    H_{l,m}^-=\frac{1}{4} \left(-
    \frac{d^2}{d\xi^2}-\frac{s(s+1)}{\cosh^2 \xi}- \frac{r(r+1)}{\cosh^2
    (\xi+i\frac{\pi}{2})}  \right)\,,
\end{equation}
where $\xi=\frac{x}{2}+i\frac{\pi}{4}$, and
\begin{equation}
    r=l+m, \qquad s=l-m-1 \,.
\end{equation}
 Representations (\ref{h1}) and (\ref{h2}) correspond, on the one hand, to
 the
 relation (\ref{T+-}) between potentials (\ref{po1}) generated by
 the shift in the half of their imaginary period,
\begin{equation}\label{imaginary}
    H_{l,m}^-(\xi)=H_{l,m}^+\left(\xi+i\frac{\pi}{2}\right) \,.
\end{equation}
On the other hand, the generators of intertwining relations
(\ref{Rlm}) correspond here to $\mathcal{R}_l:\,(r,s)\rightarrow
(-s-1,-r-1)$ and $\mathcal{R}_m:\,(r,s)\rightarrow (s,r)$.

We will return to (\ref{imaginary}) in the discussion of the
supersymmetric structure, but here we note that in the context of
periodic (elliptic) finite-gap potentials, a so called
tri-supersymmetry appears when the superpartner potentials are the
associated Lam\'e potentials shifted mutually in the half of the
real period \cite{trisusy}. As we will show below, the extended
system ${\cal H}_{l,m}$ constructed from (\ref{hamil}) also
possesses a tri-supersymmetry in which the imaginary period, the
${\cal PT}$- symmetry, the discrete symmetries (\ref{dis1}) and
(\ref{dis2}), and the non-linear supersymmetry together play a
fundamental role to form altogether a unified structure.

As a final comment on relation of (\ref{po1}) with periodic
finite-gap potentials, we note  that there is a generalization of
the family of the associated Lam\'e potentials known as the
Darboux-Treibich-Verdier potentials \cite{Veselov}. In terms of the
(double periodic) Jacobi elliptic functions  they read
\begin{eqnarray}
    V^{DTV}
     &=& n_{1}(n_{1}+1)k^{2}\mathrm{sn}^{2}x
    +n_{2}(n_{2}+1)k^{2}\mathrm{sn}^{2}(x+iK')\nonumber\\
    &+&n_{3}(n_{3}+1)k^{2}\mathrm{sn}^{2}(x+K+iK')+
    n_{4}
    (n_{4}+1)
    k^{2}\mathrm{sn}^{2}(x+K) \label{dtv}\\
    &=&
    n_{1}(n_{1}+1)k^{2}\mathrm{sn}^{2}x
    +n_{2}(n_{2}+1)\frac{1}
    {\mathrm{sn}^{2}x%
    }
    +n_{3}(n_{3}+1)\frac{\mathrm{dn}^{2}x}{\mathrm{cn}^{2}x}+
    n_{4}
    (n_{4}+1)%
        \frac{k^{ 2}
        \mathrm{cn}^{2}x}{\mathrm{dn}^{2}x}\,.\nonumber
\end{eqnarray}
Here $0<k<1$ is the modular parameter, and (\ref{dtv}) has a real,
$2K$, and an imaginary, $2iK'$, periods;  $K=K(k)$ is the elliptic
complete integral of the first kind and $K'=K(k')$,
$k'=\sqrt{1-k^2}$ \cite{specialfunctions}.  The finite-gap nature of
(\ref{dtv}) appears when parameters $n_i$ take integer values.
Particularly, when $n_2=n_3=0$, potential (\ref{dtv}) reduces to the
finite-gap associated Lam\'e potential \cite{trisusy, asso}.
 When the modular parameter takes the limit
$k\rightarrow 1$, the real period tends to infinity, $2K \rightarrow
\infty$, while $2iK'\rightarrow i\pi$, and the potential transforms
into
\begin{equation}
    V^{DTV}\xrightarrow[k\rightarrow{}1]\,
    -\frac{n_1(n_1+1)}{\cosh^2
    x}-\frac{n_2(n_2+1)}{\cosh^2 (x+i\frac{\pi}{2})}+const\,,
\end{equation}
that has the form of potentials (\ref{h1}) and (\ref{h2}).
\vskip0.1cm

 In the next section we will study the states of the
${\cal PT}$-symmetric systems $H^{\pm}_{lm}$ in the light of the
discrete symmetries and relations that we have discussed.

\section{Wavefunctions and differential intertwiners}\label{eigen}

The group theoretical methods and supersymmetry are the powerful
tools in the study of quantum mechanical systems. The Hamiltonians
(\ref{hamil}) provide a good example of the systems for which these
techniques work effectively, particularly, to analyze the spectrum
and eigenfunctions. In this direction, using irreducible
representations of the $sl(2,\mathbb{C})$ algebra, it was found in
\cite{GrNe} that the non-degenerate parts of the spectra of
potentials (\ref{po1}) are described by the two sets of
eigenfunctions, one of which is\footnote{ It is worth to note that
in the early stages of studying the complexified Scarf II potential,
just one series of the singlet states, (\ref{ps1}), that comes by
analytic continuation of the Hermitian version, was considered in
the literature \cite{susypt3, Ahmed}. Later, the complete set was
found by algebraic methods in \cite{GrNe} and reconfirmed in
refs. \cite{LeZno, LevaiCanVen2, AhmedAd}. The second, lost set of
states,
(\ref{ps2}), corresponds to resonances in the Hermitian counterpart
potential. Within the problem for the complexified Scarf II
potential this second set can be obtained from the counterparts of
the singlet states of the Hermitian problem by applying the symmetry
transformation of the parameters (\ref{dis3}).}
\begin{equation}\label{ps1}
    \Psi^{\pm}_{n,m}={\rm sech}^{m}\,  x \,\, \exp{[\mp i l\arctan (\sinh
x)]} P_{n}^{l- m-1/2,-l-m-1/2}(\pm i \sinh x) \,,
\end{equation}
where $P_{n}^{\,\alpha,\,\beta}(x)$ are the Jacobi polynomials
\cite{specialfunctions}. The corresponding energy levels and the values of the
parameter $n$ are
\begin{equation}\label{e1}
    E_{n,m}=-(m-n)^2, \quad n=0,1,2...\leq m \,.
\end{equation}
The eigenfunctions $\Psi^{\pm}_{n,m}$ for $n<m$ describe $m$ bound
states, while $n=m$ corresponds to the singlet zero energy state at
the bottom of the continuous spectrum. With the discrete symmetry
(\ref{dis1+}) of the potentials,  $V_{l,m}^\pm=V_{m+
\frac{1}{2},l-\frac{1}{2}}^\pm$, to which corresponds a
transformation
\begin{eqnarray}\label{dis3}
    &(l,m) \rightarrow \left(m+\frac{1}{2}, l-
    \frac{1}{2} \right)\,,&
\end{eqnarray}
it is possible to write down the another set of the bound states,
\begin{equation}\label{ps2}
    \Psi^{\pm}_{n,l}={\rm sech}^{l-1/2}\,  x \,
    \, \exp{[\mp i (m+1/2)\arctan (\sinh x)]}
    P_{n}^{m
    +1/2-l,-m-1/2-l}( \pm i \sinh x) \,,
\end{equation}
with energies
\begin{equation}\label{e2}
    E_{n,l}=-(l-n-1/2)^2,\quad n=0,1,2...<l-1/2 \,,
\end{equation}
so that (\ref{ps1}) and (\ref{ps2}) represent together the $l+m+1$
singlet states of the systems (\ref{hamil}).

As we will see, the separation of singlet states of each subsystem
$H^+_{l,m}$ and $H^-_{l,m}$ into the two subsets is reflected by a
specific nonlinear supersymmetry of the extended system
$\mathcal{H}_{l,m}$. This supersymmetry is related to the (imaginary
here) mutual half-period shift of the subsystems. To the best of our
knowledge a similar kind of supersymmetric structure, that we shall
discuss in the next section, was discussed till the moment only for
finite-gap systems with Hermitian Hamiltonians \cite{trisusy}. To
explain this structure and its origin, we present below  some
further comments on the properties of the Hamiltonians (\ref{hamil})
and their eigenfunctions.

States (\ref{ps1}) and (\ref{ps2}) can also be obtained from a
supersymmetry approach, by means of the Darboux-Crum transformations
(for details we refer to \cite{MatSal}), so that it is possible to link
the Hamiltonians with different values of $l$ and $m$ between them.
Particularly, it is possible to relate the free particle, for which
$H^{\pm}_{0,0}=-\frac{d^2}{dx^2}\equiv H_{0}$, with the generic case
$H^{\pm}_{l,m}$. Note that the free particle system is presented
equivalently here also by $H^{\pm}_{0,-1}$,  $H^{\pm}_{1/2,-1/2}$
and $H^{\pm}_{-1/2,-1/2}$. The bound states described above can be
computed from the appropriate non-physical states of the free
particle, whereas the states from the continuous part of the
spectrum are obtained from the plane wave states of the free system.
To illustrate this picture we will show how the scattering sector of
$H_{l,m}^{+}$ can be obtained having in mind that all the results
for $H_{l,m}^{-} $ can be reproduced then with the help of
symmetries (\ref{pt1}), or by the shift for the half of the
imaginary period. The construction of the bound states from the
non-physical states of the free particle will be illustrated in
Section \ref{examples}.

 Let us define the first order differential
operators
\begin{equation}\label{aop}
    A_{l,m}^{\pm}=\frac{d}{dx}\pm\left[ \left(l-\frac{1}{2}
    \right)\tanh x+i\left(m+
    \frac{1}{2} \right){\rm sech} \, x \right] \, .
\end{equation}
They generate the intertwining relations
\begin{eqnarray}\label{Ainter}
    &A_{l,m}^+H_{l,m}^{+}= H_{l-1,m}^{+}A_{l,m}^+ \,,\qquad
     A_{l,m}^-H_{l-1,m}^+= H_{l,m}^+A_{l,m}^- \,.&
\end{eqnarray}
Now, by applying, as in the singlet states case, the discrete
symmetry (\ref{dis3}) to (\ref{aop}), we obtain the operators
\begin{equation}\label{bop}
    B_{l,m}^{\pm}=\frac{d}{dx}\pm\left(m \tanh x+i l
    \,{\rm sech} \, x \right) \, .
\end{equation}
Application of symmetry (\ref{dis3}) to (\ref{Ainter}) results then
in the intertwining relations
\begin{eqnarray}\label{Binter}
    &B_{l,m}^+H_{l,m}^+= H_{l,m-1}^+B_{l,m}^+  \, ,\qquad
    B_{l,m}^-H_{l,m-1}^+= H_{l,m}^+B_{l,m}^- \, .&
\end{eqnarray}
The operators $A_{l,m}^{\pm}$ are related between themselves by
$A_{l,m}^{\pm}={\cal P}(A_{l,m}^{\mp})^\dagger{\cal P}^{-1}$.
Similarly, for the operators $B_{l,m}^{\pm}$ we have
$B_{l,m}^{\pm}={\cal P} (B_{l,m}^{\mp})^\dagger{\cal P}^{-1}$. Such
relations of conjugation underly  the pseudo-supersymmetry discussed
in the literature for non-Hermitian systems \cite{Mpseudosusy},
particularly,  a ${\cal PT }$-symmetric one.

Coherently with the discrete symmetry (\ref{dis3}), operators
(\ref{aop}) and (\ref{bop}) allow us to factorize, up to an additive
constant term, the same Hamiltonian in two different ways
\cite{bifurcationComment},
\begin{equation}
    H_{l,m}^{+}=-A_{l,m}^{-}A_{l,m}^{+}-
    \left(l-\frac{1}{2} \right)^2=-
    B_{l,m}^{-}B_{l,m}^{+}-m^2 \, .
\end{equation}
Since  $A_{l,m}^{\pm}$ as well as $B_{l,m}^{\pm}$ are
$\mathcal{PT}$-antisymmetric ($\mathcal{PT}$-odd),
$\{\mathcal{PT},A_{l,m}^{\pm}\}=\{\mathcal{PT},B_{l,m}^{\pm}\}=0$,
the Hamiltonians $H_{l,m}^{\pm}$ are the  $\mathcal{PT}$-symmetric
($\mathcal{PT}$-even) operators.

In the next section we will see further implications of existence of
these two related types of factorization operators. For the moment,
it is worth to note that the action of the operators $A_{l,m}^{\pm}$
and $B_{l,m}^{\pm}$ on the two-parametric family of Hamiltonians
(\ref{hamil}) is quite simple: while the former act by lowering and
raising the parameter $l$, the latter play the same role for $m$.
Using this fact, we are able now to connect, for any $l$ and $m$,
the Hamiltonian $H_{l,m}^+$ with the free particle Hamiltonian
$H_{0}$ in two different ways by making use of the operators
\begin{eqnarray}\label{freecrum1}
  {\cal D}_{l,m}^{-}&=&B_{l,m}^-B_{l,m-1}^-\ldots
    B_{l,2}^-B_{l,1}^-A_{l,0}^
    -A_{l-1,0}^-\ldots A_{2,0}^-A_{1,0}^- \, , \\
    \tilde{{\cal D}}_{l,m}^{-}&=&B_{-l,-m}^+B_{-l,-m+1}^+\ldots B_{-l,-1}^+B_{-l,0}^+A_{-l+1,0}^+A_{-l+2,0}^+\ldots A_{-1,0}^+A_{0,0}^+ \, .
    \label{freecrum2}
\end{eqnarray}
The differential operator ${\cal D}_{l,m}^{-}$ has here the order
$m+l$, meanwhile  $\tilde{\cal D}_{l,m}^{-}$ has the order $m+l+1$.
These operators intertwine Hamiltonian $H^+_{l,m}$ with the free
particle Hamiltonian,
\begin{equation}\label{HlmH0}
    {\cal D}_{l,m}^{-}H_{0}=
    H_{l,m}^+{\cal D}_{l,m}^{-}\,,
    \qquad  \tilde{{\cal D}}_{l,m}^{-}H_{0}=
    H_{l,m}^+\tilde{{\cal D}}_{l,m}^{-} \, .
\end{equation}
From (\ref{HlmH0})  the inverse intertwining relations are easily
obtained by defining ${\cal D}_{l,m}^{+}={\cal P}({\cal
D}_{l,m}^{-})^\dagger{\cal P}^{-1}$ and $\tilde{{\cal
D}}_{l,m}^{+}=-{\cal P} (\tilde{{\cal D}}_{l,m}^{-})^\dagger{\cal
P}^{-1}$, that  yields
\begin{equation}\label{freecrum3}
    {\cal D}_{l,m}^{+}H_{l,m}^+=
    H_{0}{\cal D}_{l,m}^{+}\,, \qquad
    \tilde{{\cal D}}_{l,m}^{+}H_{l,m}^+=
    H_{0}\tilde{{\cal D}}_{l,m}^{+} \, .
\end{equation}

\begin{figure}[h!]
\centering
\includegraphics[scale=1.3]{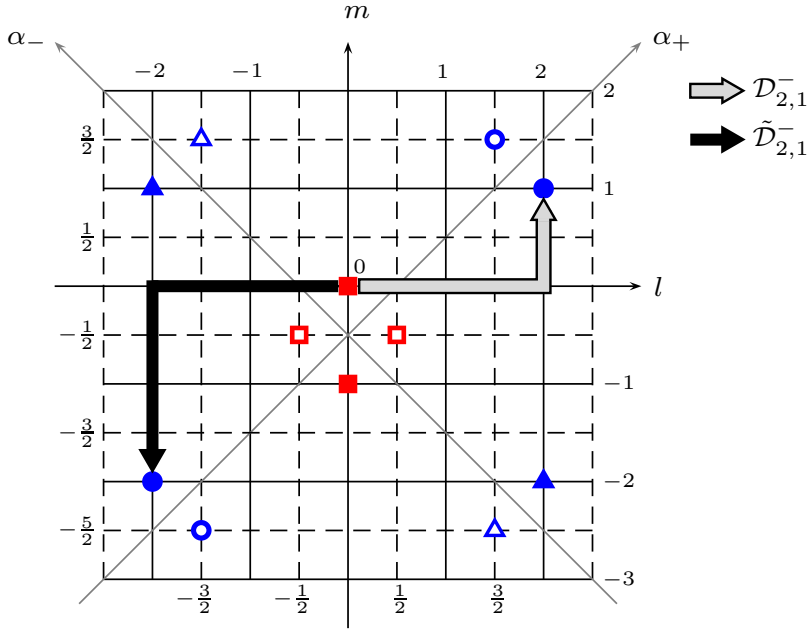}
\caption{The family of  complexified reflectionless Scarf II systems
may be presented on integer or half-integer lattices  in  the
$(l,m)$ parameters plane [in axes  $\alpha_\pm=m\pm l+\frac{1}{2}$,
the lattices for $H^\pm_{l,m}$ have half-integer coordinates]. Four
different points correspond to the same system, two of which, shown
here as an example by the filled circles, are on the integer
lattice; another two points, shown by the unfilled circles, are on
the half-integer lattice. The equivalent points on the same lattice
are related by the symmetry transformation (\ref{dis1}), while
equivalent points on different lattices are related by the discrete
symmetries (\ref{dis1+}) and (\ref{dis2}). The filled and unfilled
triangles correspond to a system shifted in the half of the pure
imaginary period; two such mutually displaced systems are related by
the intertwining discrete transformations (\ref{pt1}) and
(\ref{Rlm}). The two filled and two unfilled squares correspond  to
a free particle system. Any two systems represented on the same
lattice may be related between themselves by differential
intertwining operators. Particularly, any nontrivial complexified
reflectionless Scarf II system may be intertwined with the free
particle. Two of such (of many possible) "intertwining paths" shown
for the system $H_{2,1}^+=H^+_{-2,-2}$ correspond to the action of
the operators (\ref{freecrum1}) and (\ref{freecrum2}) with $l=2$,
$m=1$.} \label{red}
\end{figure}

Several comments are in order here. First we note that the
difference of orders of the two intertwining operators
(\ref{freecrum1}) and of their conjugate ones are related to the
fact that the system $H^+_{l,m}$ can be presented alternatively by
the equivalent Hamiltonian $H^+_{l,-m-1}$. Another point is worth to
note is that since operators $A_{l,m}^{\pm}$ and $B_{l,m}^{\pm}$ do
not commute, and the path that connects the points $(l,m)$ and
$(0,0)$ in the parameter plane can be chosen in different ways, the
corresponding intertwining operators have not a unique form. It is
the operators ${\cal D}_{l,m}^{\pm}$ and $\tilde{\cal
D}_{l,m}^{\pm}$ that together with the corresponding pair of
Hamiltonians $H^\pm_{l,m}$ will form a basis for the construction of
the tri-supersymmetric structure in the present
$\mathcal{PT}$-symmetric case, see Figure \ref{red}.

On the other hand, the fact that two different Darboux-Crum
transformations can relate two different quantum mechanical systems
is known for the case of Hermitian operators and was exploited in
\cite{trisusy} to reveal a peculiar,  tri-supersymmetric structure in some
periodic and non-periodic finite-gap systems.  Particularly, it was
shown in \cite{newself1} that the two non-trivial Darboux-Crum
transformations encode the existence of a Lax pair in reflectionless
P\"oschl-Teller systems. As we noted above on  Eq. (\ref{ptrela}),
there are cases in which the potentials (\ref{po1}) are reduced
exactly to the shifted Hermitian P\"oschl-Teller potentials, so that
operators (\ref{freecrum1}) and (\ref{freecrum2}) match in those
particular cases the corresponding operators in \cite{trisusy, ptads}.

One of the direct  applications of the constructed intertwining
operators is that we can use them  to map the plane wave states of
the free particle,
\begin{equation}\label{pw}
    H_{0}\psi_{\pm k}=k^2\psi_{\pm k},\qquad
    \psi_{\pm k}=e^{\pm i k
    x}\,,
\end{equation}
into the scattering  eigenstates of $H_{l,m}^+$,
\begin{equation}\label{conti}
    H_{l,m}^+\Psi^{+}_{\pm k}=k^2\Psi^{+}_{\pm k},
    \qquad \Psi^{+}_{\pm
    k}={\cal D}_{l,m}^{-}e^{\pm i k x}=
    c_1(k) \tilde{{\cal D}}_{l,m}^{-}e^{\pm i k
    x}\,,
\end{equation}
where $k\geq 0$ and $c_1(k)$ is some  ($k$-dependent) constant
factor. For $k=0$, the action of the operator ${\cal D}_{l,m}^{-}$
produces  the unique singlet state of the continuous spectrum of
$H_{l,m}^+$, and then (\ref{conti}) coincides with the eigenfunction
(\ref{ps1}) with $n=m$, i.e. $\Psi^{+}_{0}=\Psi^{+}_{n,n}$. On the
other hand, $\tilde{{\cal D}}_{l,m}^{-}$ annihilates the singlet
state located at the bottom of the continuos spectrum of the free
particle, i.e. $c_1(0)=0$ in (\ref{conti}). \vskip0.1cm

This picture of the Darboux-Crum transformations explains, as in the
case of Hermitian finite-gap systems  \cite{trisusy}, the reflectionless
properties of the Hamiltonians (\ref{hamil}) for integer (and
half-integer) values of $l$ and $m$.

\section{Tri-supersymmetric structure}\label{susysec}

In this section we will show how the ${\cal PT}$-symmetry,
originated from the intertwining relations (\ref{pt1}), the discrete
symmetries behind (\ref{Rlm}), the self-isospectrality based on
(\ref{T+-}), the Lax integrals ${\mathbb A}^\pm_{2n+1}$, and the
non-linear supersymmetry form altogether a peculiar structure.

To reveal and describe such an unusual extended nonlinear
supersymmetric structure, we will show first that the extended
$\mathcal{PT}$-symmetric Hamiltonian
\begin{equation}\label{sH}
    {\cal H}_{l,m}=\left( \begin{array}{cc}
    H^{+}_{l,m} & 0  \\
    0 & H^{-}_{l,m}
\end{array}\right) \,
\end{equation}
for  $l\neq0$ has three \emph{mutually commuting} non-trivial
basic \emph{integrals} of motion, anti-diagonal ${\cal X}_{l,m}$
and ${\cal Y}_{l,m}$,  and diagonal ${\cal Z}_{l,m}=diag\,
({\mathbb A}^+_{2n+1},{\mathbb A}^-_{2n+1})$, $n=l+m$,
\begin{equation}\label{com}
    [ {\cal X}_{l,m}, {\cal H}_{l,m}]=0, \quad [ {\cal Y}_{l,m},
     {\cal H}_{l,m}]=0, \quad [ {\cal Z}_{l,m},
     {\cal H}_{l,m}]=0 \, .
\end{equation}
These are the matrix non-linear differential operators of the
orders $|{\cal X}_{l,m}|=2l$, $|{\cal Y}_{l,m}|=2m+1$ and $|{\cal
Z}_{l,m}|=2(l+m)+1$, connected between themselves by a
factorization relation
\begin{equation}\label{XYZ}
    {\cal Z}_{l,m}={\cal X}_{l,m}{\cal Y}_{l,m}=
    {\cal Y}_{l,m}{\cal X}_{l,m}\,.
\end{equation}
It is due to these three basic nontrivial integrals the
corresponding supersymmetric structure is referred to as a
\emph{tri-supersymmetry}. We will see that it reflects coherently
the peculiar properties of the extended complexified Scarf II system
$\mathcal{H}_{l,m}$, including the existence of two types of the
discrete energy levels in its spectrum.

The case $l= 0$ is particular since the extended  Hamiltonians
(\ref{sH}) just reduce to the two copies of the Hermitian
 P\"oschl-Teller systems with coupling parameter
 $\lambda=m$ in (\ref{PT}), i.e.
 $H^{+}_{0,\lambda}=H^{-}_{0,\lambda}=H^{PT}_m$.
 One integral then reduces to the Pauli sigma matrix, ${\cal
 X}_{0,m}=\sigma_1$.
 The remaining two integrals are related in accordance
 with (\ref{XYZ}) as  ${\cal Y}_{0,m}=\sigma_1{\cal Z}_{0,m}$,
 where the diagonal matrix elements of ${\cal Z}_{0,m}$ coincide,
 ${\mathbb A}^+_{2m+1}={\mathbb A}^-_{2m+1}$,
 and generate the hidden bosonized nonlinear
 supersymmetry of the reflectionless P\"oschl-Teller
 system $H^{PT}_m$, see \cite{boso1}.

Before passing  over to the construction of the nontrivial
integrals, we note that the extended system (\ref{sH}) is formed by
the two self-isospectral Hamiltonians displaced  mutually by the
half of their imaginary period. As a result, the degeneracy of the
spectrum of $ {\cal H}_{l,m}$ is twice that of its corresponding
diagonal components.
  As in usual (Hermitian) quantum mechanics, by virtue of relations
  (\ref{com}),
  it is natural to expect that there is a basis where all the eigenstates
  of the Hamiltonian (\ref{sH}) are also the eigenstates of  the nontrivial
  integrals of motion  $\mathcal{X}_{l,m}$, $\mathcal{Y}_{l,m}$, and $\mathcal{Z}_{l,m}$.
 In accordance with this, as we will see,
  the $m+l$ doublet states corresponding to the set of bound states
  can be presented in the form
\begin{equation}\label{phi1}
 \Phi_{n,l}^{\pm}=\left( \begin{array}{c}
    \Psi_{n,l}^+   \\
   \pm i \Psi_{n,l}^-
    \end{array}\right), \qquad n=0,1,\ldots, <l-1/2\,,
\end{equation}
and
\begin{equation}\label{phi2}
 \Phi_{n,m}^{\pm}=\left( \begin{array}{c}
    \Psi_{m,l}^+   \\
   \pm \Psi_{m,l}^-
    \end{array}\right), \qquad n=0,1,2...< m\,,
\end{equation}
with eigenvalues given by (\ref{e2}) and (\ref{e1}), respectively.
The  scattering states can be written as
\begin{equation}\label{contspectr}
 \Phi^{\pm}_{+k}=
 \left( \begin{array}{c}
    \Psi_{+k}^+   \\
   \pm \Psi_{+k}^-
    \end{array}\right), \qquad  \Phi^{\pm}_{-k}=
 \left( \begin{array}{c}
    \Psi_{-k}^+   \\
   \pm \Psi_{-k}^-
    \end{array}\right),
\end{equation}
with energies $E=k^2$, $k\geq 0$. The energy levels with $E>0$  are
then four-fold degenerate, while for $E=0$ we  have
$\Phi^{\pm}_{0}=\Phi^{\pm}_{+k}=\Phi^{\pm}_{-k}$,
 and, as in the bound states case, the double degeneration.
 Up to a multiplicative constant,
 $\Phi^{\pm}_{0}$ coincide with (\ref{phi2}) with $n=m$.

The intertwining relations (\ref{Rlm}) for the potentials are
trivially extended for the Hamiltonians (\ref{hamil}),
\begin{equation}\label{comrlm}
    \mathcal{R}_{l}H^{\pm}_{l,m}=
    H^{\mp}_{l,m}\mathcal{R}_{l},  \qquad
    \mathcal{R}_{m}H^{\pm}_{l,m}=
    H^{\mp}_{l,m}\mathcal{R}_{m}\,,
\end{equation}
where the generators $\mathcal{R}_{l}$ and $\mathcal{R}_{m}$  can be
used to construct a discrete symmetry
$\mathcal{R}_{l}\mathcal{R}_{m}$ for the extended Hamiltonian ${\cal
H}_{l,m}$. For the extended system ${\cal H}_{l,m}$, antidiagonal
Pauli matrix $\sigma_1$ (as well as $\sigma_2$), produces the same
effect of intertwining of the Hamiltonian's components, $\sigma_1
diag\,(H^{+}_{l,m},H^{-}_{l,m})\sigma_1=
diag\,(H^{-}_{l,m},H^{+}_{l,m})$. Therefore, we can construct  the
matrix operators
\begin{equation}
    \hat{\mathcal{R}}_l=\sigma_1\mathcal{R}_l, \qquad
    \hat{\mathcal{R}}_m=\sigma_1\mathcal{R}_m \, ,
\end{equation}
which are  the integrals of motion for our extended system
(\ref{sH}),
\begin{equation}
    [\hat{\mathcal{R}}_l,{\cal H}_{l,m}]=0,  \qquad
    [\hat{\mathcal{R}}_m,{\cal H}_{l,m}]=0 \, .
\end{equation}
In the case of $\hat{\mathcal{R}}_l$, the  commutation relation
comes from the intertwining relation, which, in turn, is based on
the equality  $H^{\pm}_{- l,m}=H^{\mp}_{l,m}$. In the previous
section we have seen that the operators $A^{\pm}_{l,m}$ acting on
the Hamiltonans $H^{\pm}_{l,m}$, can lower or raise the index $l$,
see Eq. (\ref{Ainter}), by means of a chain of Darboux
transformations. This means that the appropriate product of the
operators $A^{\pm}_{l,m}$ produces exactly the same intertwining
effect as the ${\cal R}_l$, which changes $(l,m)$ for $(-l,m)$.
Indeed, this can be achieved by application of the $2l$-th order
differential operators
\begin{eqnarray}\label{xdef}
    X_{l,m}^+&\equiv& A_{-l+1,m}^+A_{-l+2,m}^+ \ldots A_{0,m}^+ \ldots
    A_{l-1,m}^+A_{l,m}^+ \,, \\
    X_{l,m}^-&\equiv& A_{l,m}^- A_{l-1,m}^-\ldots A_{0,m}^- \ldots A_{-l
    +2,m}^- A_{-l+1,m}^-\,,
\end{eqnarray}
which satisfy the intertwining relations of the same form as in
(\ref{comrlm}),
\begin{equation}\label{in1*}
    X_{l,m}^{\pm}H_{l,m}^{\pm}=H_{-l,m}^{\pm}X_{l,m}^{\pm}=H_{l,m}^{
    \mp}X_{l,m}^{\pm} \, .
\end{equation}
The case $l=0$  reduces trivially to the operators
$X_{0,m}^{\pm}=1$. On the other hand, the nontrivial analog of the
discrete symmetry $\hat{{\cal R}}_l$, $\hat{{\cal R}}_l^2=1$, is
provided by  the matrix differential operator,
\begin{equation}\label{xdef*}
    {\cal X}_{l,m}=\left( \begin{array}{cc}
    0 &X_{l,m}^-  \\
    X_{l,m}^+ & 0
    \end{array}\right) \,,
\end{equation}
which, by virtue of (\ref{in1*}), is an integral of motion. In
addition to  (\ref{com}), the integral (\ref{xdef*}) satisfies a
superalgebraic-type relation
\begin{equation}\label{xh*}
  \left \{  {\cal X}_{l,m},{\cal
    X}_{l,m}
    \right \}=2{\cal X}_{l,m}^2=2P_{{\cal X}}({\cal H}_{l,m})\,,
\end{equation}
where $P_{{\cal X}}$ is a polynomial of order $l$ in Hamiltonian
${\cal H}_{l,m}$,
\begin{equation}\label{px}
    P_{{\cal X}}({\cal H}_{l,m})=\prod^{[l-1/2]}_{s=0}({\cal
    H}_{l,m}+(l- s-1/2)^2)^2 \,.
\end{equation}
In the context of analogy of ${\cal X}_{l,m}$ with $\hat{{\cal
R}}_l$, Eq.  (\ref{xh*}) is a  generalization  of the relation
$\hat{{\cal R}}_l^2=1$. The polynomial $P_{{\cal X}}$  has the
nature of a spectral polynomial, but which only includes the
energies of one set of bound states (\ref{phi1}). {}From here a
remarkable property of ${\cal X}_{l,m}$ can be derived:  acting on
eigenstates of ${\cal H}_{l,m}$, it annihilates one complete set of
doublets while the doublet states of another set  are the
eigenvectors with nonzero eigenvalues,
\begin{equation}\label{xphi}
 {\cal X}_{l,m}\Phi_{n,l}^{\pm}=0, \quad  {\cal X}_{l,m}
 \Phi_{n,m}^{\pm}=\pm (-1)^n\prod^{[l-1/2]}_{s=0}(E_{n,m}+
    (l-s-1/2)^2) \Phi_{n,m}^{\pm}\,,
\end{equation}
for $ n=0,1,\ldots, <l-1/2$ and $ n=0,1,2,\ldots\leq m$. Therefore,
the integral ${\cal X}_{l,m}$ identifies all the states which
correspond to resonances of the Hermitian counterparts of the
Hamiltonians $H^\pm_{l,m}$ \cite{LevaiCanVen2}. The action of the operator
${\cal X}_{l,m}$ on the scattering states is characterized by the
property that it does not distinguish the waves coming from the left
or from the right, but separates  the states with distinct values of
the upper index,
\begin{eqnarray}
 {\cal X}_{l,m}^{\pm}\Phi^{\pm}_{+k}&=& \pm (-1)^l
 \prod^{[l-1/2]}_{s=0}(k^2+(l-
    s-1/2)^2)\Phi^{\pm}_{+k}\,,\nonumber\\
     {\cal X}_{l,m}^{\pm}\Phi^{\pm}_{-k}&=& \pm  (-1)^l\prod^{[l-1/2]}_{s=0}
     (k^2+(l-
    s-1/2)^2)\Phi^{\pm}_{-k} \, .\label{Xcont}
\end{eqnarray}

We can construct also differential operators of order $2m+1$,
\begin{eqnarray}\label{ydef}
    Y_{l,m}^+&\equiv& B_{l,-m}^+B_{l,-m+1}^+ \ldots B_{l,0}^+ \ldots
    B_{l,m-1}^+B_{l,m}^+ \,, \\
    Y_{l,m}^-&\equiv& B_{l,m}^-B_{l,m-1}^- \ldots B_{l,0}^- \ldots B_{l,-m+1}^-B_{l,-m}^-\,,
\end{eqnarray}
which generate the Darboux-Crum transformations similar to the
intertwining relations produced by $\mathcal{R}_m$,
\begin{equation}\label{iny1}
    Y_{l,m}^{\pm}H_{l,m}^{\pm}=
    H_{l,-m-1}^{\pm}Y_{l,m}^{\pm}=H_{l,m}^{\mp}Y_{l,m}^{\pm} \,.
\end{equation}
With their help, we find that the matrix differential operator
\begin{equation}\label{Yint}
    {\cal Y}_{l,m}=i\left( \begin{array}{cc}
    0 &Y_{l,m}^-  \\
    Y_{l,m}^+ & 0
    \end{array}\right) \,
\end{equation}
is the another nontrivial integral of motion for the extended system
$\mathcal{H}_{l,m}$. Like the $\hat{\mathcal{R}}_{m}$ commutes with
the  $\hat{\mathcal{R}}_l$, the nontrivial integrals (\ref{Yint})
and  (\ref{xdef*}) also commute,
\begin{equation}
    [{\cal X}_{l,m},{\cal Y}_{l,m}]=0 \, .
\end{equation}
The integral ${\cal Y}_{l,m}$ generates a relation
\begin{equation}\label{py}
    \left \{  {\cal Y}_{l,m},{\cal
    Y}_{l,m}
    \right \}=2P_{{\cal Y}}({\cal H}_{l,m})=
    2{\cal H}_{l,m}\prod^{m-1}_{r=0}({\cal H}_{l,m}+
    (m-r)^2)^2
\end{equation}
to be of the form similar to (\ref{xh*}). The roots of the spectral
polynomial $P_{{\cal Y}}({\cal H}_{l,m})$ are complementary to those
of the polynomial $P_{{\cal X}}({\cal H}_{l,m})$: they coincide with
the energies (\ref{e1}) of the eigenstates (\ref{phi2}). In
correspondence with this property, the second non-trivial integral
of motion, ${\cal Y}_{l,m}$, annihilates the remaining set of
discrete eigenstates of $\mathcal{H}_{l,m}$, not annihilated by the
integral $\mathcal{X}_{l, m}$,  which correspond to the doubly
degenerate energy levels, while the zero modes of the latter
integral are the eigenstates of ${\cal Y}_{l,m}$ of the nonzero
eigenvalues,
\begin{equation}\label{yphi}
     {\cal Y}_{l,m}\Phi_{n,l}^{\pm}=\pm i(-1)^n |E_{n,l}|^{1/2}
     \prod^{m-1}_{r=0}(E_{n,l}+(m-
    r)^2)\Phi_{n,l}^{\pm} ,
    \quad  {\cal Y}_{l,m}\Phi_{n,m}^{\pm}=0 \, .
\end{equation}
The appearance of imaginary eigenvalues in the spectrum of the
integral ${\cal Y}_{l,m}$  will be discussed later. The action of
${\cal Y}_{l,m}$ on the states of the continuous spectrum
(\ref{contspectr}) is given by
\begin{eqnarray}\label{Yc1}
 {\cal Y}_{l,m}\Phi^{\pm}_{+k}&=& \mp (-1)^m k\prod^{m-1}_{r=0}(k^2+(m-r)^2) \Phi^{\pm}_{+k}\, ,\\
{\cal Y}_{l,m}\Phi^{\pm}_{-k}&=& \pm (-1)^m k\prod^{m-1}_{r=0}(k^2+(m-r)^2) \Phi^{\pm}_{-k} \, ,
\label{Yc2}
\end{eqnarray}
i.e. this integral, unlike the ${\cal X}_{l,m}$, see (\ref{Xcont}),
distinguishes the waves coming from the left and from the right, and
as ${\cal X}_{l,m}$, detects a difference between the states with
distinct values of the upper (sign) index.

As we have seen, behind the existence of the integrals of motion $
{\cal X}_{l,m}$ and $ {\cal Y}_{l,m}$ is the fact that there are two
\emph{different} Darboux-Crum transformations, which intertwine the
Hamiltonians $H^+_{lm}$ and $H^-_{l,m}$. In the case of the discrete
operators ${\cal R}_l$ and ${\cal R}_m$ (one can also  consider the
intertwining operators ${\cal P}$,  ${\cal T}$ and $T_{\pm}$, see
the discussion in Section \ref{sec1}), their composition transforms
into a symmetry operation for the Hamiltonians $H^\pm_{l,m}$,
\begin{equation}\label{rlmh}
[{\cal R}_l{\cal R}_m, H^{\pm}_{l,m}]=0\,.
\end{equation}
For extended  system (\ref{sH}), this composition corresponds to the
integral  $\hat{{\cal R}}=\hat{{\cal R}}_l\hat{{\cal
R}}_m=diag({\cal R}_l{\cal R}_m,{\cal R}_l{\cal R}_m)$,
\begin{equation}
    [\hat{\mathcal{R}}, {\cal H}_{l,m}]=0\,.
\end{equation}
The composition of the intertwining relations  generated by
$X^{\pm}_{l,m}$ and $Y^{\mp}_{l,m}$, (\ref{in1*}) and (\ref{iny1})
respectively, yields
\begin{equation}
    (Y_{l,m}^{\mp}X_{l,m}^{\pm})H_{l,m}^{\pm}=H_{l,m}^{\pm}
    (Y_{l,m}^{\mp}X_{l,m}^{\pm}), \qquad
    (X_{l,m}^{\mp}Y_{l,m}^{\pm})H_{l,m}^{\pm}=
    H_{l,m}^{\pm}(X_{l,m}^{\mp}Y_{l,m}^{\pm})
    \, .
\end{equation}
The intertwining relations transform therefore into commutation
relations, and corresponding  integral of motion appears for each
Hamiltonian, analogously to (\ref{rlmh}). The resulting integrals
are the differential operators of the order $2n+1$ with $n=m+l$, and
these are nothing else as the Lax integrals ${\mathbb A}^\pm_{2n+1}$
in (\ref{Adef}), which we rename here as
\begin{equation}
    Z^{\pm}_{l,m}={\mathbb A}^{\pm}_{2n+1}=Y_{l,m}^{\mp}X_{l,m}^{\pm}=
    X_{l,m}^{\mp}Y_{l,m}^{\pm}, \qquad [Z^{\pm}_{l,m},H_{l,m}^{\pm}]=0
    \,.
\end{equation}
For the extended system, these integrals of motion can be joined to
form  a diagonal operator, ${\cal Z}_{l,m}$, which is generated by
the anticommutator of the previous conserved quantities,
\begin{equation}\label{zdef}
    {\cal Z}_{l,m}=i\left( \begin{array}{cc}
    Z_{l,m}^+ & 0 \\
     0 & Z_{l,m}^-
    \end{array}\right)=\frac{1}{2}\{{\cal X}_{l,m},{\cal Y}_{l,m} \}
    \,.
\end{equation}

\begin{figure}[h!]
\centering
\includegraphics[scale=1.3]{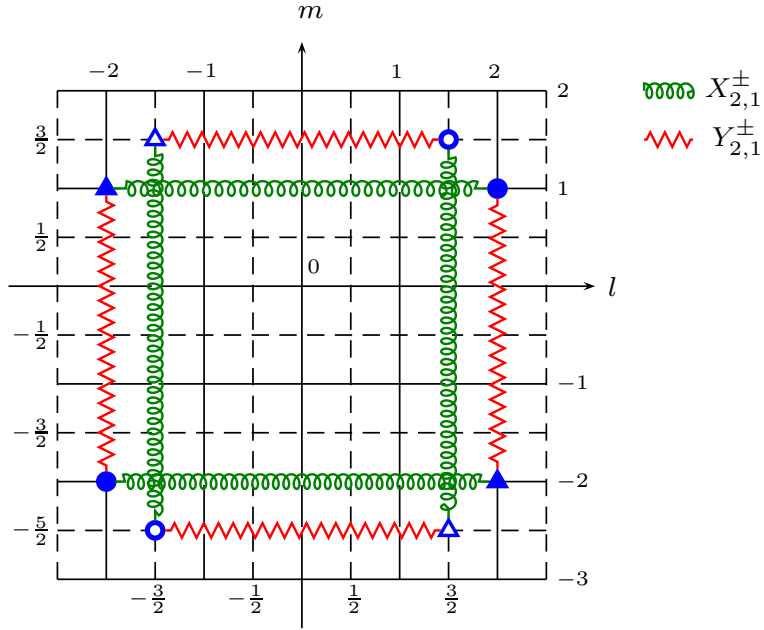}
\caption{The ``intertwining paths", which correspond to the action
of the operators $X_{2,1}^{\pm}$ and $Y_{2,1}^{\pm}$ between the
mutually shifted, self-isospectral  Hamiltonians (presented by
circles and triangles), are shown for the case of the extended
$\mathcal{PT}$-symmetric system
$\mathcal{H}_{2,1}=\mathcal{H}_{\frac{3}{2},\frac{3}{2}}$. The
effect produced by these intertwining operators on the
integer-valued lattice is the same as that of the discrete operators
${\cal R}_l$ and ${\cal R}_m$, respectively. On the
half-integer-valued lattice, the horizontal and vertical distances
between the corresponding systems are interchanged, and in this
sense, the action of the operators $X_{2,1}^{\pm}$ and
$Y_{2,1}^{\pm}$ on the half-integer-valued lattice is dual to that
on the integer-valued lattice.  Starting from any point, the
composition of horizontal and vertical paths to the equivalent point
on the same lattice produces the Lax integrals $Z_{2,1}^{\pm}$ and
the symmetry operator ${\cal R}_l{\cal R}_m$. } \label{red5}
\end{figure}

The origin of the Lax integrals in the present extended Hamiltonian
from the the intertwining operators $X_{l,m}^{\pm}$ and
$Y_{l,m}^{\pm}$ is illustrated on Fig. {\ref{red5}}

It is straightforward to check from the above relations that  ${\cal
Z}_{l,m}$ commutes with the Hamiltonian (\ref{com}) and with the
integrals ${\cal X}_{l,m}$ and ${\cal Y}_{l,m}$,
\begin{equation}
    [{\cal Z}_{l,m},{\cal X}_{l,m}]=0,
    \qquad [{\cal Z}_{l,m},{\cal Y}_{l,m}]=0 \, .
\end{equation}
Its square produces a polynomial in ${\cal H}_{l,m}$,
\begin{equation}\label{pz}
    {\cal P}_{{\cal Z}}({\cal H}_{l,m})={\cal H}_{l,m}\prod^{m-1}_{r=0}
    \prod^{[l-1/2]}_{s=0}({\cal H}_{l,m}+(m-r)^2)^2({\cal
    H}_{l,m}+(l-s-1/2)^2)^2 \,.
\end{equation}
whose roots are all the discrete doubly degenerate energies of the
extended system. Note that the roots of the bound states are of
degree two, while the zero energy state root has degree one. All the
corresponding energy eigenstates are the zero modes of the integral
${\cal Z}_{l,m}$,
\begin{equation}\label{zphi}
     {\cal Z}_{l,m}\Phi_{n,l}^{\pm}=0,
     \qquad  {\cal Z}_{l,m}\Phi_{n,m}^{\pm}=0,
\end{equation}
which also detects  the direction of propagation of the waves of the
scattering sector,
\begin{eqnarray}
 {\cal Z}_{l,m}\Phi^{\pm}_{+k}&=& -(-1)^{m+l}
 k\prod^{m-1}_{r=0}\prod^{[l-1/2]}_{s=0}(k^2+(m-r)^2)(k^2+(l-
    s-1/2)^2)\Phi^{\pm}_{+k}\,,\\
     {\cal Z}_{l,m}\Phi^{\pm}_{-k}
     &=& (-1)^{m+l} k\prod^{m-1}_{r=0}
     \prod^{[l-1/2]}_{s=0}((k^2+(m-r)^2)k^2+(l-
    s-1/2)^2)\Phi^{\pm}_{-k} \, .
\end{eqnarray}

Because of the $\mathcal{PT}$-odd nature of the operators
$A^\pm_{l,m}$ and  $B^\pm_{l,m}$, from which ${\cal X}_{l,m}$,
${\cal Y}_{l,m}$ and $ {\cal Z}_{l,m}$ are composed, all this
triplet of the integrals  is $\mathcal{PT}$-even. Thus, instead to
be Hermitian operators, all the conserved quantities commute with
the $ {\cal PT}$ operator,
\begin{equation}\label{compt}
    [{\cal H}_{l,m},{\cal PT}]=0, \quad
    [{\cal X}_{l,m},{\cal PT}]=0, \quad [{\cal
    Y}_{l,m},{\cal PT}]=0, \quad
    [{\cal Z}_{l,m},{\cal PT}]=0 \, .
\end{equation}
It is this property of the $\mathcal{PT}$-symmetry that requires the
presence of the imaginary unit as a multiplicative factor in the
definition of ${\cal Y}_{l,m}$ in
Eq. (\ref{Yint}).  In turn, the factor $i$ in Eq. (\ref{Yint})
emphasizes then the existence of the splitting of the discrete
eigenstates into two different families, and reveals an additional
specific feature of the whole supersymmetric configuration we have
here. The simultaneous requirement of a common basis of eigenstates
for all the integrals of motion in addition to the relation
(\ref{compt}) fixes that only the integral $ {\cal Y}_{l,m}$ has
imaginary eigenvalues for the set of doublet states $
\Phi_{n,l}^{\pm}$ in Eq. (\ref{phi1}) (see Eq. (\ref{yphi}) and also
the examples in the next section). As all the integrals are mutually
commuting operators, the picture  here is different from that in a
usual quantum mechanics where Hermitian (self-adjoint) mutually
commuting operators possess a common basis of eigenstates with real
eigenvalues. It is instructive to look in more detail what happens
here.  Explicit form of the states in (\ref{phi1}) shows  that they
are not the  eigenstates of the ${\cal PT}$ operator. Indeed, they
satisfies the relation
\begin{equation}
    {\cal PT}\Phi_{n,l}^{\pm}=
    \Phi_{n,l}^{\mp} \, .
\end{equation}
Remembering that ${\cal X}_{l,m}$ and ${\cal Z}_{l,m}$ annihilate
the set of doublet states (\ref{phi1}), while the extended
Hamiltonian (\ref{sH})  has an entire real spectrum [and the states
(\ref{phi1}) are its eigenstates], one concludes that the ${\cal
PT}$-symmetry has a \emph{broken} nature just for the integral
${\cal Y}_{l,m}$ [we remind  parenthetically here that the
eigenstates from the continuous part of the spectrum have real
eigenvalues for ${\cal Y}_{l,m}$, see Eqs. (\ref{Yc1}) and
(\ref{Yc2})]. Taking into account independently only the Hamiltonian
operator, one can find another basis where these states are
simultaneously the eigenstates of the Hamiltonian and the ${\cal
PT}$ operator;  therefore,  for the ${\cal H}_{l,m}$ the ${\cal
PT}$-symmetry is \emph{unbroken}.

We have identified the nontrivial integrals of the extended system
and discussed their properties. Now we consider the related
nonlinear supersymmetric structure. The diagonal matrix $\sigma_3$
is a trivial integral of motion for $\mathcal{H}_{l,m}$.
Nevertheless it allows us to double the set of the nontrivial
integrals of motion since the multiplication of any of them by
$\sigma_3$ gives a new, linear independent nontrivial matrix
integral of motion. So, in this way we obtain the set of six
linearly independent nontrivial matrix integrals of motion
\begin{eqnarray}
&\mathcal{Q}^{(1)}_{l,m}= {\cal X}_{l,m}\,,\qquad
\mathcal{Q}^{(2)}_{l,m}= \sigma_3\mathcal{Q}^{(1)}_{l,m}\,,&\label{QQ}\\
&\mathcal{S}^{(1)}_{l,m}= {\cal Y}_{l,m}\,,\qquad
\mathcal{S}^{(2)}_{l,m}=\sigma_3\mathcal{S}^{(1)}_{l,m}\,,\label{SS}&\\
&\mathcal{L}^{(1)}=\mathcal{Z}_{l,m}\,,\qquad
\mathcal{L}^{(2)}=\sigma_3\mathcal{L}^{(1)}_{l,m}\,.\label{LL}&
\end{eqnarray}
Notice the absence of the imaginary factor $i$ in the definition of
the second anti-diagonal supercharges in comparison with the usual
SUSYQM approach with a  Hermitian Hamiltonian. This guarantees that
all the three new integrals are also ${\cal PT}$-symmetric
operators.

The square of the matrix integral  $\sigma_3$ equals $1$, and it can
be identified as the grading operator, $\Gamma=\sigma_3$. This
grading operator classifies then the Hamiltonian $\mathcal{H}_{l,m}$
and integrals $\mathcal{L}^{(a)}_{l,m}$, $a=1,2$, as bosonic
operators, while the integrals of the antidiagonal matrix form,
$\mathcal{Q}^{(a)}_{l,m}$ and $\mathcal{S}^{(a)}_{l,m}$, are
classified as fermionic operators. In correspondence with this, we
get a nonlinear superalgebra with the following set of nontrivial
(anti)-commutation relations:
\begin{eqnarray}
    &\{\mathcal{Q}^{(a)}_{l,m},
    \mathcal{Q}^{(b)}_{l,m}\}=(-1)^{a+1}2\delta_{ab}
    P_{{\cal X}},\,\,\{\mathcal{S}^{(a)}_{l,m},\mathcal{S}^{(b)}_{l,m}\}
    =(-1)^{a+1}2\delta_{ab}P_{{\cal Y}},\,\,
    \{ \mathcal{Q}^{(a)}_{l,m}, \mathcal{S}^{(b)}_{l,m} \}=-2
    \delta_{ab} \mathcal{L}^{(1)}_{l,m}
    \, , &\nonumber \\
    & [\mathcal{Q}^{(a)}_{l,m},\mathcal{L}^{(2)}_{l,m}]
    =(-1)^{a}2\epsilon_{ab} \mathcal{S}^{(b)}_{l,m}P_{{\cal X}},
    \quad [\mathcal{S}^{(a)}_{l,m}, \mathcal{L}^{(2)}_{l,m}]
    =(-1)^{a}2\epsilon_{ab} \mathcal{Q}^{(b)}_{l,m}
    P_{{\cal Y}} \, , & \label{susyalg}
\end{eqnarray}
where $P_{{\cal X}}=P_{{\cal X}}(\mathcal{H}_{l,m})$ and $P_{{\cal
Y}}=P_{{\cal Y}}(\mathcal{H}_{l,m})$  are the polynomials defined in
(\ref{px}), (\ref{py}) and (\ref{pz}), respectively. Note that the
integral $\mathcal{L}^{(1)}_{l,m}$ commutes with all the other
integrals and, so,  plays here the role of the bosonic central
charge.

The choice of $\sigma_3$  as the grading operator is, however, not
unique. Another possibility corresponds, for instance, to the choice
$\Gamma=\mathcal{P}T_+$ (or, $\Gamma=\mathcal{P}T_-$). Indeed, this
operator is a (nonlocal) integral of motion, whose square is equal
to  $1$. Such a grading operator classifies the integrals
$\mathcal{Q}^{(a)}_{l,m}$ as bosonic integrals, while
$\mathcal{S}^{(a)}_{l,m}$ and $\mathcal{L}^{(a)}_{l,m}$ are
classified as fermionic integrals. The corresponding superalgebraic
relations can be computed then by making use of the relations
described above. In this case we have, particularly,  a relation
$\{\mathcal{L}^{(a)}_{l,m},\mathcal{L}^{(b)}_{l,m}\}=2\delta_{ab}
P_{{\cal Z}}(\mathcal{H}_{l,m})$. This corresponds to the fact that
each of the  unextended $\mathcal{PT}$-symmetric systems $H^+_{l,m}$
and $H^-_{l,m}$ is characterized by the  bosonized supersymmetry, in
which the $\mathcal{PT}$-symmetric integrals  $iZ^+_{l,m}$ and
$iZ^-_{l,m}$, respectively,   are treated as the
$\Gamma=\mathcal{P}T_+$-odd supercharges.

\begin{figure}[h!]
\centering
\includegraphics[scale=0.9]{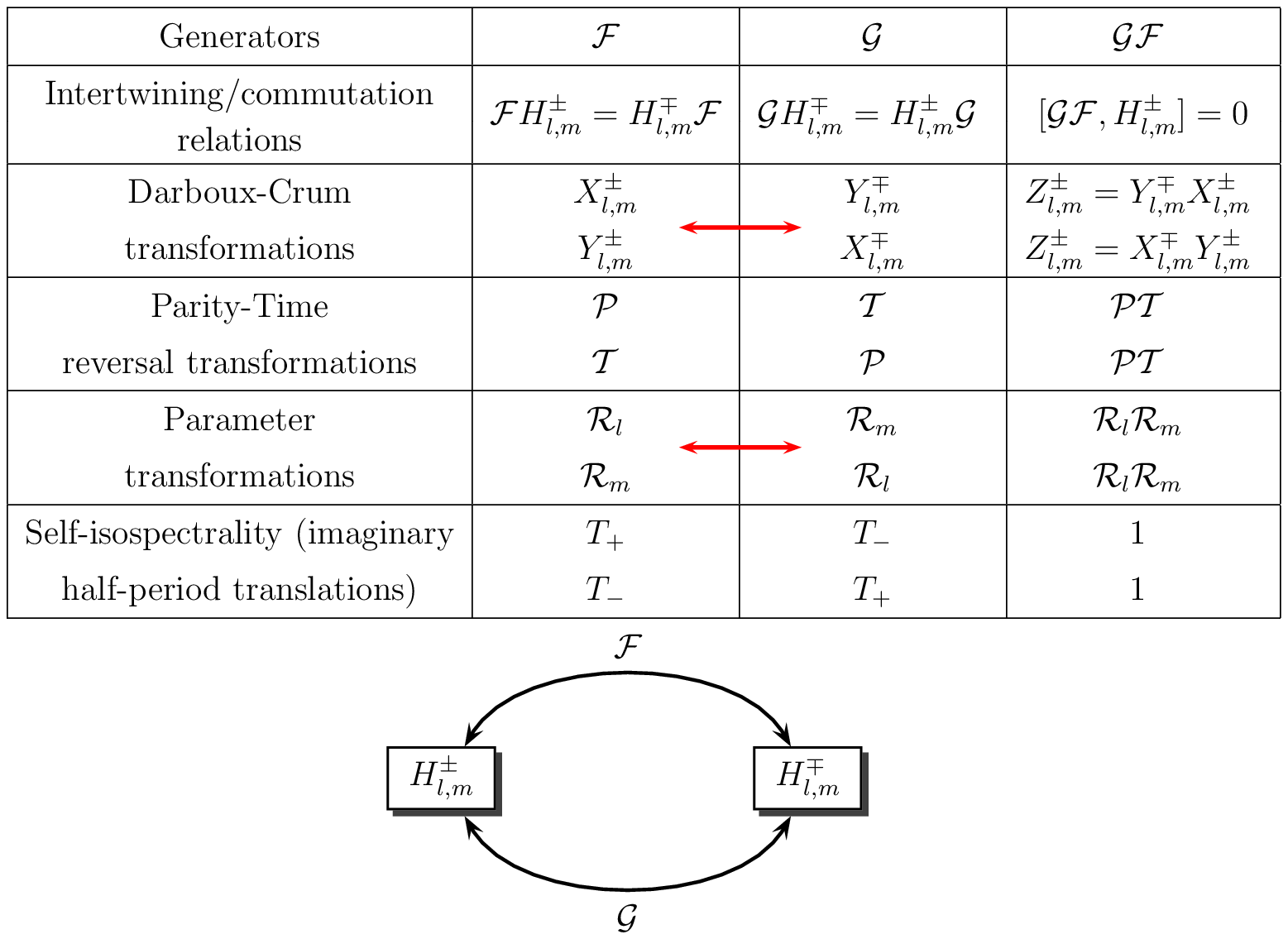}
\caption{The basic blocks of the tri-supersymmetric structure of the
extended $\mathcal{PT}$-symmetric Hamiltonian $\mathcal{H}_{l,m}$.
Horizontal arrows inside a table correspond to the duality
transformation induced by the symmetry $(l,m) \rightarrow
\left(m+\frac{1}{2}, l-
    \frac{1}{2} \right)$.}\label{tablita}
\end{figure}

We summarize the whole picture on which  the tri-supersymmetric
structure is based on Fig. \ref{tablita}.

\section{Examples}\label{examples}

To illustrate different properties of the systems $H^\pm_{l,m}$ and
the tri-supersymmetric structure of $\mathcal{H}_{l,m}$, here we
present some examples for specific values of $l$ and $m$. Before
doing this, we first note that according to the relation
(\ref{ptrela}), the simplest nontrivial case  of
$H^+_{1,0}(x)=H^-_{1,0}(x+i\pi)$ reduces, up to rescaling, just to
the displaced reflectionless P\"oschl-Teller system with one bound
state in the spectrum. The unique bound state corresponds to a
resonance with a complex energy value in the spectrum of the
Hermitian Hamiltonian with real Scarf II potential $V(x)={\rm
sech}^2(x)-\sinh x\,{\rm sech}^2x$, which is depicted on Fig.
\ref{Scarf}. So we will consider more rich cases of reflectionless
$\mathcal{PT}$-symmetric systems with two and three bound states.

\subsection{Systems with two bound states}

Without loss of generality, the family of reflectionless potentials
$V^\pm_{l,m}$ with two bound states in the discrete part of the
spectrum can be presented by non-negative integer values of the
parameters $l$ and $m$ subjected to the condition $l+m=2$. The case
$ (l=0,m=2)$ corresponds here to the well known Hermitian
reflectionless P\"oschl-Teller potential $V_{0,2}^+=V_{0,2}^-
=-6/\cosh^2 x$. By virtue of (\ref{ptrela}), this potential shares,
up to rescaling, the same spectrum  as complexified Scarf II
potentials with $ (l=1,m=1)$. Then the remaining case $(l=2,m=0)$,
\begin{equation}\label{h20}
    H^{\pm}_{2,0}=
    -\frac{d^2}{dx^2}-\frac{4}{\cosh^2x}\pm2 i\frac{ \sinh
    x} {\cosh^2x} \,,
\end{equation}
provides a first nontrivial example which is not related to a
Hermitian (reflectionless P\"oschl-Teller) counterpart by means of
shifting and rescaling of the coordinate. The potentials
$V=V_{2,0}^{\pm}$ are solutions of the nonlinear $s$-KdV$_2$
equation
\begin{equation}
    V^{(5)}-10V'''V-20V'V''+30V^2V'-10(V'''-3V V')+9V'=0 \,,
\end{equation}
where $V^{(5)}=d\,^5V/dx^5$. Notice that in
contrast with the complexified case $V_{2,0}^{\pm}$, the Hermitian
version of the potential, $V_{2i,0}$, which is plotted on Fig.
\ref{Scarf}, is not a solution of the $s$-KdV$_2$ equation. The
Hamiltonians $H^{\pm}_{2,0}$ fall into the class of systems studied
in \cite{Wadati} in the context of the $\mathcal{PT}-$symmetric
nonlinear integrable systems; the corresponding extended Hamiltonian
$\mathcal{H}_{2,0}$ appears as a particular case of the diagonalized
squared Dirac equation for a free spin-1/2 field in de Sitter space,
see Ref. \cite{Maloney}.

The degeneracy of the spectrum of ${\cal H}_{2,0}$ is twice  that
for each system in (\ref{h20}), and we have two doublets of  bound
states and two zero energy states at the very bottom of the
four-fold degenerate continuous part of the spectrum. The bound
states correspond here to the complex energy resonances in the
Hermitian version with real Scarf II potential,  see Fig. 2. Their
eigenfunctions,
\begin{equation}
    \Phi_{0,l}^{\pm}=\left( \begin{array}{c}
    \displaystyle \frac{e^{- \frac{i}{2}\arctan
    \sinh x}}{\cosh^{3/2}x}   \\
    \\    \displaystyle \pm i \frac{e^{  \frac{i}{2}\arctan
    \sinh x}}{\cosh^{3/2}x}
    \end{array}\right), \quad \Phi_{1,l}^{\pm}=\left( \begin{array}{c}
      \displaystyle  \frac{e^{ -\frac{i}{2}\arctan \sinh x}}{\cosh^{3/2}x}(1-
    2i\sinh x) \\ \\
   \displaystyle  \pm i \frac{e^{\frac{i}{2}\arctan \sinh x}}{\cosh^{3/2}x}(1+
    2i\sinh x)
    \end{array}\right), \label{phi0}
\end{equation}
satisfy equations
\begin{equation}\label{h20ener}
    {\cal H}_{2,0}\Phi_{0,l}^{\pm}=
    -\frac{9}{4}\Phi_{0,l}^{\pm}, \qquad {\cal
    H}_{2,0} \Phi_{1,l}^{\pm}=
    -\frac{1}{4} \Phi_{1,l}^{\pm} \,,
\end{equation}
and correspond to wavefunctions (\ref{phi1}).

 These solutions can be obtained from the non-physical states of
the free particle, by applicacion of the operators (\ref{freecrum1})
or (\ref{freecrum2}) with $l=2$ and $m=0$ in the case of the
subsystem $H^{+}_{2,0}$. For generic values of $l$ and $m$, the
Darboux-Crum transformations that map the free particle eigenstates
into those  for the lower Hamiltonian $H^{-}_{l,m}$ are realized in
correspondence with intertwining relations (\ref{freecrum3}) by
means of the operators
\begin{eqnarray}\label{freecrum3}
  {\cal D}_{l,m}^{\sharp}&=& {\cal P}
  ({\cal D}_{l,m}^{-}){\cal P} ={\cal T}
  ({\cal D}_{l,m}^{-}){\cal T} \, , \\
    \tilde{{\cal D}}_{l,m}^{\sharp}&=& -{\cal P}
    (\tilde{{\cal D}}_{l,m}^{-}){\cal
    P} =-{\cal T} (\tilde{{\cal D}}_{l,m}^{-}){\cal T}\,.
    \label{freecrum4}
\end{eqnarray}
Note that in correspondence with relation
$H^{-}_{l,m}(x)=H^{+}_{l,m}(x+i\pi)$, operators (\ref{freecrum3})
are obtained equivalently from the intertwining operators ${\cal
D}_{l,m}^{-}$ and $\tilde{{\cal D}}_{l,m}^{-}$ by the half-period
shift.  The non-physical states which are transformed into bound
states by means of the Darboux-Crum transformations
(\ref{freecrum1}), (\ref{freecrum2}), (\ref{freecrum3}) and
(\ref{freecrum4}) are
\begin{equation}\label{9/4}
    \phi_{-9/4}^{+}=\cosh \left( \frac{3x}{2}
    \right), \qquad \phi_{-9/4}^{-}=\sinh
    \left( \frac{3x}{2} \right) \, ,
\end{equation}
and
\begin{equation}\label{1/4}
    \phi_{-1/4}^{+}=\cosh
    \left( \frac{x}{2} \right), \qquad \phi_{-1/4}^{-}=\sinh
    \left( \frac{x}{2} \right) \, ,
\end{equation}
which obey the same Schr\"odinger equations as in (\ref{h20ener}),
\begin{eqnarray}\label{nonphysicalenergies}
    H_{0}\, \phi_{-9/4}^{\pm}=
    -\frac{9}{4}\phi_{-9/4}^{\pm}, \quad H_{0}\,
    \phi_{-1/4}^{\pm}=-\frac{1}{4}\phi_{-1/4}^{\pm} \, .
\end{eqnarray}
One can choose solutions of different parity with respect to ${\cal
P}$ in (\ref{9/4}) and (\ref{1/4}) to obtain the bound states
(\ref{phi0}). Choosing the functions with positive ${\cal P}$-parity
we have,
\begin{eqnarray}
    \Phi_{0,l}^{\pm}&=&\frac{2}{3}\left( \begin{array}{c}
   {\cal D}_{2,0}^{-}
    \\   \pm i\,  {\cal D}_{2,0}^{\sharp}
    \end{array}\right) \phi_{-9/4}^{+}=\frac{4i}{9}
    \left( \begin{array}{c}
    \tilde{{\cal D}}_{2,0}^{-}
    \\   \mp i\,   \tilde{{\cal D}}_{2,0}^{\sharp}
    \end{array}\right) \phi_{-9/4}^{+} \, , \label{r1} \\
    \Phi_{1,l}^{\pm}&=&-2\left( \begin{array}{c}
    {\cal D}_{2,0}^{-}
    \\   \pm i\,  {\cal D}_{2,0}^{\sharp}
    \end{array}\right) \phi_{-1/4}^{+}=4i\left( \begin{array}{c}
   \tilde{{\cal D}}_{2,0}^{-}
    \\   \mp i\,   \tilde{{\cal D}}_{2,0}^{\sharp}
    \end{array}\right) \phi_{-1/4}^{+} \,, \label{r2}
\end{eqnarray}
where
    ${\cal D}_{2,0}^{-}=
    A^-_{2,0}A^-_{1,0}$, $\tilde{{\cal
    D}}_{2,0}^{-}=B^+_{-2,0}A^+_{-1,0}A^+_{0,0}$,
    ${\cal D}_{2,0}^{\sharp}=
    A^-_{0,-1}A^-_{-1,1}$, $\tilde{{\cal
    D}}_{2,0}^{\sharp}=
    B^+_{2,0}A^-_{2,0}A^-_{1,0}$.
Expressions (\ref{r1}) and (\ref{r2}) with the non-physical states
of ${\cal P}$-negative parity, i.e. $\phi_{-9/4}^{-}$ and
$\phi_{-1/4}^{-}$, remain almost identical up to multiplicative
constant factors. One can choose these factors pure  imaginary to
produce, for each entry of (\ref{r1}) and (\ref{r2}), a state of
definite ${\cal PT}$-parity by starting from $\phi_{-9/4}^{\pm}$ or
$\phi_{-1/4}^{\pm}$. This can be understood by taking into account
that the intertwining operators have no definite ${\cal P}$-parity,
but they have a definite ${\cal PT}$-parity. In general case, while
the ${\cal D}_{l,m}^{-}$ and ${\cal D}_{l,m}^{\sharp}$ are ${\cal
PT}$-even, the operators $\tilde{{\cal D}}_{l,m}^{-}$ and
$\tilde{{\cal D}}_{l,m}^{\sharp}$ are ${\cal PT}$-odd.

Using the same procedure as with bound states, we can construct the
eigenstates in  the scattering sector (\ref{contspectr}) by applying
the intertwining Darboux-Crum operators to the plane waves
(\ref{pw}) of $H_{0}$,
\begin{equation}\label{sca20}
    \Phi_{+k}^{\pm}=\left( \begin{array}{c}
   {\cal D}_{2,0}^{-}
    \\   \pm \,  {\cal D}_{2,0}^{\sharp}
    \end{array}\right)e^{ikx}, \qquad \Phi_{-k}^{\pm}=\left( \begin{array}{c}
   {\cal D}_{2,0}^{-}
    \\   \pm \,  {\cal D}_{2,0}^{\sharp}
    \end{array}\right)e^{-ikx}\,,
\end{equation}
\begin{equation}
    {\cal H}_{2,0}\Phi_{+k}^{\pm}=
    k^2\Phi_{+k}^{\pm}\,,\qquad {\cal
    H}_{2,0}\Phi_{-k}^{\pm}=k^2\Phi_{-k}^{\pm} \, .
\end{equation}
The eigenfunctions (\ref{sca20}) for $H_{2,0}^+$ and $H_{2,0}^-$ may
also be obtained by applying, instead, the operators $\tilde{{\cal
D}}_{2,0}^{-}$ and $\tilde{{\cal D}}_{2,0}^{\sharp}$, respectively,
to the same plane wave eigenstates. In the case of the zero energy
eigenstates we have
\begin{equation}
    \Phi_{0}^{\pm}=
    \left( \begin{array}{c}
   {\cal D}_{2,0}^{-}
    \\   \pm \,  {\cal D}_{2,0}^{\sharp}
    \end{array}\right)1=\left( \begin{array}{c}
    \tilde{{\cal D}}_{2,0}^{-}
    \\   \pm \,  \tilde{{\cal D}}_{2,0}^{\sharp}
    \end{array}\right)x=-\frac{3}{4}\left( \begin{array}{c}
    e^{-2i\arctan \sinh x }  \\
    \pm e^{2i\arctan \sinh x }
    \end{array}\right),
\end{equation}
i. e. the operators $\tilde{{\cal D}}_{2,0}^{-}$ and $\tilde{{\cal
D}}_{2,0}^{\sharp}$ should act on the non-physical zero energy
solutions of $H_{0}$ which are proportional to $x$. Another solution
of zero energy of $H_{0}$, which is a constant, is annihilated by
the Darboux-Crum operators,
\begin{equation}\label{annihi}
    \tilde{{\cal D}}_{2,0}^{-} 1=
    \tilde{{\cal D}}_{2,0}^{\sharp} 1=0 \, .
\end{equation}

The extended Hamiltonian ${\cal H}_{2,0}$ possesses three basic
conserved quantities in the form of the  matrix differential
operators. One of these integrals, ${\cal X}_{2,0}$, is given by
\begin{equation}\label{x20}
    {\cal X}_{2,0}=\left( \begin{array}{cc}
    0 &X_{2,0}^-  \\
    X_{2,0}^+ & 0
    \end{array}\right)=\left( \begin{array}{cc}
    0 & A_{2,0}^-A_{1,0}^-A_{0,0}^-A_{-1,0}^-   \\
    A_{-1,0}^+A_{0,0}^+A_{1,0}^+A_{2,0}^+ & 0
    \end{array}\right) \,.
\end{equation}
The explicit form of the higher order differential operators is
\begin{eqnarray}
    &X^\pm_{2,0}=\frac{d^4}{dx^4}\pm
    \frac{2i}{\cosh x}\frac{d^3}
    {dx^3}+\frac{1}{\cosh^2 x}\left(6\mp3i\sinh x-
    \frac{5}{2}\cosh^2x  \right)
    \frac{d^2}{dx^2}&\\
    &-\frac{1}{\cosh^3 x}\left(12\sinh x
    \pm i\left[\frac{1}{2}\cosh^2 x-3\right]
    \right)\frac{d}{dx}-\frac{13}{\cosh^4x}
    \left(13 \mp 3 i\sinh x\right)\left(1\pm i
    \sinh x\right)^3\,.&  \notag
\end{eqnarray}
This integral acts on the physical states of the Hamiltonian ${\cal
H}_{2,0}$ as follows,
\begin{equation}
    {\cal X}_{2,0}\Phi_{0,l}^{\pm}=0,
    \quad {\cal X}_{2,0}\Phi_{1,l}^{\pm}=0,
    \quad
    {\cal X}_{2,0}\Phi_{0}^{\pm}=
    \pm \frac{9}{16}\Phi_{0}^{\pm}, \quad
\end{equation}
\begin{equation}
    {\cal X}_{2,0}\Phi_{+k}^{\pm}=
    \pm \left(k^2+\frac{1}{4}\right)\left(k^2+
    \frac{9}{4}\right)\Phi_{+k}^{\pm},
    \quad {\cal X}_{2,0}\Phi_{-k}^{\pm}=\pm
    \left(k^2+\frac{1}{4}\right)
    \left(k^2+\frac{9}{4}\right)\Phi_{-k}^{\pm}\,.
\end{equation}
The operator ${\cal X}_{2,0}$ does not distinguish the waves coming
from the left or the right, but recognizes the states which
correspond to resonances in the Hermitian Scarf II potential
spectrum, by annihilating all of them. Another  integral of motion
is also an anti-diagonal matrix differential operator,
\begin{equation}
    {\cal Y}_{2,0}=i\left( \begin{array}{cc}
    0 &Y_{2,0}^-  \\
    Y_{2,0}^+ & 0
    \end{array}\right)=i\left( \begin{array}{cc}
    0 & B_{2,0}^-   \\
    B_{2,0}^+ & 0
    \end{array}\right) \,,
\end{equation}
where
\begin{equation}
    Y_{2,0}^\pm=B_{2,0}^\pm=\frac{d}{dx} \pm i\frac{2}{\cosh x} \,.
\end{equation}
The integral of motion ${\cal Y}_{2,0}$ commutes with ${\cal
X}_{2,0}$ and  encodes the information to be complementary to that
provided by the latter. This can be seen from its action on the
physical states,
\begin{equation}
    {\cal Y}_{2,0}\Phi_{0,l}^{\pm}=
    \pm\frac{3i}{2}\Phi_{0,l}^{\pm}, \qquad {\cal
    Y}_{2,0}\Phi_{1,l}^{\pm}=
    \pm\frac{i}{2}\Phi_{1,l}^{\pm}, \qquad {\cal
    Y}_{2,0}\Phi_{k}^{\pm}=0, \quad
\end{equation}
\begin{equation}
    {\cal Y}_{2,0}\Phi_{+k}^{\pm}=
    \mp k \Phi_{+k}^{\pm}, \qquad {\cal
    Y}_{2,0}\Phi_{-k}^{\pm}=\pm k\Phi_{-k}^{\pm}\,.
\end{equation}
The remaining doublet states are annihilated by ${\cal Y}_{2,0}$,
which in this case correspond to the states of the zero energy at
the bottom of the continuous spectrum. The waves coming from the
left or the right are recognized by it, and like (\ref{x20}), the
${\cal Y}_{2,0}$ detects also the upper index of eigenstates. Note
that the bound eigenstates here that correspond to resonances in the
Hermitian Scarf II systems spectra have pure imaginary eigenvalues
of ${\cal Y}_{2,0}$.

Coherently with the properties of the displayed antidiagonal
integrals, the diagonal operator $ {\cal
Z}_{2,0}=\mathcal{X}_{2,0}\mathcal{Y}_{2,0}$ annihilates all the
doublet states, and separates scattering states coming from
different directions,
\begin{equation}
    {\cal Z}_{2,0}\Phi_{0,l}^{\pm}=0,
    \qquad {\cal Z}_{2,0}\Phi_{1,l}^{\pm}=0,
    \qquad
    {\cal Z}_{2,0}\Phi_{0}^{\pm}=0\,,
\end{equation}
\begin{equation}
    {\cal Z}_{2,0}\Phi_{+k}^{\pm}=
    - k\left(k^2+\frac{1}{4}\right)\left(k^2+
    \frac{9}{4}\right)\Phi_{+k}^{\pm},
    \quad {\cal Z}_{2,0}\Phi_{-k}^{\pm}= k
    \left(k^2+\frac{1}{4}\right)
    \left(k^2+\frac{9}{4}\right)\Phi_{-k}^{\pm}\,.
\end{equation}

\subsubsection{Systems with three bound states}

Reflectionless systems with three bound states are constrained to
fullfill the relation $l+m=3$. Potentials (\ref{po1}) with
$(l=0,m=3)$ and $ (l=2,m=1)$ are related by Eq. (\ref{ptrela}).
When $(l= 3,m=0)$, the systems $H^\pm_{3,0}$ have three bound
states, all of which correspond to resonance states in the
Hermitian version. So, all
the mentioned  three bound states systems have corresponding
analogs in more simple cases we have discussed above. A more
non-trivial example with three bound states is given by  the
Hamiltonians
\begin{equation}
    H^{\pm}_{1,2}=-\frac{d^2}{dx^2}
    -\frac{7}{\cosh^2x}\pm5 i\frac{ \sinh x}
    {\cosh^2x}\,.
\end{equation}
The Hermitian counterpart potential for
$H^{+}_{1,2}$ is shown on Fig. 2. Here the potentials
$V=V^{\pm}_{1,2}$ satisfy the $s$-KdV$_3$ equation
\begin{eqnarray}
    &_{}&V^{(7)}-14V(V^{(5)}-20V''V')-
    42V^{(4)}V'-70V'''V''+70V'''V^2+70V'^3-140V'V
    ^3+ \notag\\
    &_{}&-21(V^{(5)}-10V'''V-20V'V''+
    30V^2V')+84(V'''-6V V')-64V'=0\,.
\end{eqnarray}
The extended Hamiltonian ${\cal H}_{1,2}$ has six bound states,
where four correspond to the bound state from the set (\ref{ps1}),
\begin{equation}
    \Phi_{0,m}^{\pm}=\left( \begin{array}{c}
     \displaystyle \frac{e^{ - i\arctan \sinh x}}{\cosh^2 x}  \\ \\
    \pm  \displaystyle \frac{e^{ i\arctan \sinh x}}{\cosh^2 x}
    \end{array}\right), \quad \Phi_{1,m}^{\pm}=\left( \begin{array}{c}
      \displaystyle \frac{e^{ \mp i\arctan \sinh x}}{\cosh^2 x}(2-3i\sinh x)  \\ \\
     \displaystyle   \pm  \frac{e^{ \mp i\arctan \sinh x}}{\cosh^2 x}(2+3i\sinh
     x)
    \end{array}\right),
    \end{equation}
\begin{equation}\label{energies12}
    {\cal H}_{1,2}\Phi_{0,m}^{\pm}=
    -4\Phi_{0,m}^{\pm}, \qquad {\cal
    H}_{1,2}\Phi_{1,m}^{\pm}=-\Phi_{1,m}^{\pm}\,.
\end{equation}
These solutions are analogs of the bound states for the Hermitian
counterpart system, which, in addition, admits resonances for a
complex value of energy. Those resonances correspond to two extra
bound states solutions in the spectrum of the Hamiltonian ${\cal
H}_{1,2}$,
\begin{equation}
 \Phi_{0,l}^{\pm}=\left( \begin{array}{c}
    \displaystyle   \frac{e^{ -\frac{5i}{2}
    \arctan \sinh x}}{\cosh^{1/2} x}  \\ \\
     \displaystyle   \pm i \frac{e^{
     \frac{5i}{2}\arctan \sinh x}}{\cosh^{1/2} x}
     \end{array}\right)\, ,\qquad
     {\cal H}_{1,2} \Phi_{0,l}^{\pm}=-
    \frac{1}{4} \Phi_{0,l}^{\pm} \, .
\end{equation}
All the bound states described above, $\Phi_{0,m}$, $\Phi_{1,m}$ and
$\Phi_{0,l} $, can be derived from the non-physical states of the
free particle, $ \phi_{-4}^{\pm}$, $\phi_{-1}^{\pm}$ and
$\phi_{-1/4}^{\pm}$, respectively. Here, the unphysical solutions of
$H_{0}$ are
\begin{equation}\label{4}
    \phi_{-4}^{+}=\cosh 2x ,
    \qquad \phi_{-4}^{-}=\sinh 2x  \, ,
\end{equation}
and
\begin{equation}\label{1}
    \phi_{-1}^{+}=\cosh x ,
    \qquad \phi_{-1}^{-}=\sinh x\,;
\end{equation}
they have the same eigenvalues (of $H_0$) as the bound states
$\Phi_{0,m}^{\pm}$ and $ \Phi_{1,ml}^{\pm}$ in (\ref{energies12}).
The mapping between the states is given by
\begin{eqnarray}
    &\Phi_{0,m}^{\pm}=\frac{2i}{15}
    \left( \begin{array}{c}
   {\cal D}_{1,2}^{-}
    \\   \mp \,  {\cal D}_{1,2}^{\sharp}
    \end{array}\right) \phi_{-4}^{+} , \quad \Phi_{1,m}^{\pm}=-\frac{2}
    {3}\left( \begin{array}{c}
   {\cal D}_{1,2}^{-}
    \\   \pm \,  {\cal D}_{1,2}^{\sharp}
    \end{array}\right) \phi_{-1}^{+}, \quad \Phi_{0,l}^{\pm}=-\frac{8}
    {15}\left( \begin{array}{c}
   {\cal D}_{1,2}^{-}
    \\   \pm i \,  {\cal D}_{1,2}^{\sharp}
    \end{array}\right) \phi_{-1/4}^{-}\,,& \nonumber
\end{eqnarray}
where we use the definition of the operators (\ref{freecrum1}) and
(\ref{freecrum3}). Similar expressions can be found by making use of
the operators (\ref{freecrum2}) and (\ref{freecrum4}); it is worth
to note, however, that some non-physical states of the free particle
cannot be mapped properly because are annihilated,
\begin{eqnarray}
     {\cal D}_{1,2}^{-} \phi_{-4}^{-}=
     {\cal D}_{1,2}^{\sharp} \phi_{-4}^{-}=
     {\tilde{\cal D}}_{1,2}^{-} \phi_{-4}^{+}
     ={\tilde{\cal D}}_{1,2}^{\sharp}
    \phi_{-4}^{+}=0 \, , \\
    {\cal D}_{1,2}^{+} \phi_{-1}^{-}=
    {\cal D}_{1,2}^{\sharp} \phi_{-1}^{+}=
    {\tilde{\cal D}}_{1,2}^{-} \phi_{-1}^{-}
    ={\tilde{\cal D}}_{1,2}^{\sharp}
    \phi_{-1}^{-}=0 \, .
\end{eqnarray}
The situation is quite similar to the previous case (\ref{annihi}). In fact, in
this case the constant state of the free particle is also annihilated by the
 operators ${\tilde{\cal D}}_{1,2}^{-}1={\tilde{\cal
 D}}_{1,2}^{\sharp}1=0$.

The wave functions of the continuum are obtained by the same method
from the free plane waves,
\begin{equation} \label{sca12}
    \Phi_{+k}^{\pm}=\left( \begin{array}{c}
    {\cal D}_{1,2}^- \\
    \pm {\cal D}_{1,2}^\sharp
    \end{array}\right)e^{ikx}, \qquad \Phi_{-k}^{\pm}=
    \left( \begin{array}{c}
    {\cal D}_{1,2}^- \\
    \pm {\cal D}_{1,2}^\sharp
    \end{array}\right)e^{-ikx} \, ,
\end{equation}
and have energies $E=k^2$.

The anti-diagonal,  mutually commuting basic integrals of motion for
this case, ${\cal X}_{1,2}$ and ${\cal Y}_{1,2}$, have differential
orders $|{\cal X}_{1,2}|=2$ and $|{\cal Y}_{1,2}| =5$, and read
\begin{equation}\label{x12}
    {\cal X}_{1,2}=\left( \begin{array}{cc}
    0 &X_{1,2}^-  \\
    X_{1,2}^+  & 0
    \end{array}\right)=\left( \begin{array}{cc}
    0 &A_{1,2}^-A_{2,0}^-  \\
    A_{2,0}^+A_{1,2}^+  & 0
    \end{array}\right) \,,
\end{equation}
 and
\begin{equation}\label{y12}
    {\cal Y}_{1,2}=i\left( \begin{array}{cc}
    0 &Y_{1,2}^-  \\
    Y_{1,2}^+  & 0
    \end{array}\right)=i\left( \begin{array}{cc}
    0 &  B_{1,2}^-B_{1,1}^-B_{1,0}^-B_{1,-1}^-B_{1,-2}^-  \\
   B_{1,-2}^+B_{1,-1}^+B_{1,0}^+B_{1,1}^+B_{1,2}^+  & 0
    \end{array}\right) \,.
\end{equation}
The explicit form of differential operators that compose (\ref{x12})
and (\ref{y12}) are
\begin{equation}
    X_{1,2}^\pm=\frac{d^2}{dx^2}\pm
    \frac{5i}{\cosh x}\frac{d}{dx}-\frac{1}{8
    \cosh^2 x}\left(44+\cosh^2x\pm 20i \sinh x \right)
\end{equation}
and
\begin{eqnarray}
    &Y_{1,2}^\pm=\frac{d^5}{dx^5}\pm
    \frac{5i}{\cosh x}\frac{d^4}{dx^4}-
    \frac{5}{\cosh^2x}\left( \sinh^2 x\pm
    2i  \sinh x\right)\frac{d^3}{dx^3}-
    \frac{5}{\cosh^3x}\left( 3\sinh x\mp2i
    \pm i\sinh^2 x \right)\frac{d^2}{dx^2} &
    \notag \\&+
    \frac{1}{\cosh^4 x}\left(4\cosh^4 x+10\cosh^2x-30\mp65i\sinh x\pm10i\sinh
    x \right)\frac{d}{dx}& \nonumber \\&
    -\frac{15}{\cosh^5 x}\left(2\sinh x\mp5i
    \pm4i\cosh^2 x \right) \nonumber \,.&
\end{eqnarray}

The action of the integrals on the doublets of the Hamiltonian is
given by
\begin{equation}
    {\cal X}_{1,2}\Phi_{0,m}^{\pm}=
    \mp \frac{15}{4}\Phi_{0,m}^{\pm}, \quad
    {\cal
    X}_{1,2}\Phi_{1,m}^{\pm}=
    \pm \frac{3}{4}\Phi_{1,m}^{\pm},  \quad {\cal
    X}_{1,2}\Phi_{0,l}^{\pm}=0,
    \quad {\cal X}_{1,2}\Phi_{0,l}^{\pm}=\pm
    \frac{1}
    {4}\Psi_{k,0}^{\pm},
\end{equation}
\begin{equation}
    {\cal Y}_{1,2}\Phi_{0,m}^{\pm}=0,
    \quad {\cal Y}_{1,2}\Phi_{1,m}^{\pm}=0,
    \quad
    {\cal Y}_{1,2}\Phi_{0,l}^{\pm}=
    \pm \frac{45i}{32}\Phi_{0,l}^{\pm}, \quad
    {\cal
    Y}_{1,2}\Phi_{0}^{\pm}=0 \, .
\end{equation}
Note that, again, the states $\Phi_{0,l}^{\pm}$,  which correspond
to resonances in the Hermitian counterpart systems, are annihilated
by the integral $\mathcal{X}_{1,2}$ and are characterized by pure
imaginary eigenvalues of the second integral $\mathcal{Y}_{1,2}$.
The scattering states (\ref{sca12}) are eigenstates of the operators
${\cal X}_{1,2}$ and ${\cal Y}_{1,2}$,
\begin{equation}
    {\cal X}_{1,2}\Phi_{+k}^{\pm}=
    \mp \left(k^2+\frac{1}{4}\right)
    \Phi_{+k}^{\pm}, \qquad {\cal X}_{1,2}
    \Phi_{-k}^{\pm}=\mp \left(k^2+\frac{1}
    {4}\right)\Phi_{-k}^{\pm}\,,
\end{equation}
\begin{equation}
    {\cal Y}_{1,2}\Phi_{+k}^{\pm}=
    \mp k(k^2+1)(k^2+4)\Phi_{+k}^{\pm}, \qquad
    {\cal Y}_{1,2}\Phi_{-k}^{\pm}=
    \pm k(k^2+1)(k^2+4) \Phi_{-k}^{\pm}\,.
\end{equation}
Finally, the diagonal integral ${\cal Z}_{1,2}={\cal Y}_{1,2}{\cal
X}_{1,2}={\cal X}_{1,2}{\cal Y}_{1,2}$ annihilates the whole set of
doublet states,
\begin{equation}
    {\cal Z}_{1,2}\Phi_{0,m}^{\pm}=0,
    \quad {\cal Z}_{1,2}\Phi_{1,m}^{\pm}=0,
    \quad
    {\cal Z}_{1,2}\Phi_{0,l}^{\pm}=0, \quad {\cal
    Z}_{1,2}\Phi_{0}^{\pm}=0 \, ,
\end{equation}
and recognizes, as the integral $ {\cal Y}_{1,2}$, the waves coming
from the left or the right,
\begin{equation}
    {\cal Z}_{1,2}\Phi_{+k}^{\pm}=
    k\left(k^2+\frac{1}{4}\right)(k^2+1)(k^2+4)
    \Phi_{+k}^{\pm}, \quad {\cal Z}_{1,2}
    \Phi_{-k}^{\pm}= -k\left(k^2+\frac{1}
    {4}\right)(k^2+1)(k^2+4) \Phi_{-k}^{\pm}\,.
\end{equation}

\section{Discussion and outlook}\label{discus}

In this paper, by analyzing a two-parametric family of
reflectionless $ {\cal PT}$-symmetric Hamiltonians, we have revealed
a new supersymmetric structure. The class of potentials studied here
provides an instructive example of quantum mechanical systems with
non-Hermitian Hamiltonians. In comparison with the Hermitian version
of the Scarf II potential, the spectrum of its complexified
counterpart contains two series of singlet states discovered earlier
within a framework of the group theoretical approach. Surprisingly,
this characteristic is imprinted in a tri-supersymmetric structure
that is based here on the specific properties of the family of
potentials: their pure imaginary period and discrete symmetries of a
reflection type in the indexes. Usually, the imaginary period in
both Hermitian and non-Hermitian Hamiltonians does not play
explicitly an important role at the level of the spectrum, or in
supersymmetric aspects. Following the original idea of Dunne and
Feinberg for the case of a usual SUSYQM with a linear Lie
superalgebraic structure and mutually shifted (on a real line)
Hermitian Hamiltonians \cite{Dunne}, we construct an extended
$\mathcal{PT}$-symmetric system
composed by two Hamiltonians with self-isospectral potentials, but
now displaced mutually in the half of the \emph{imaginary} period.
The obtained composed system has three basic non-trivial integrals
of motion,  which in the generic case are the higher order
differential operators. The importance of the splitting of the
discrete states becomes clear by analyzing these integrals. Two of
the anti-diagonal, supercharge-type integrals, $ {\cal X}_{l,m}$ and
${\cal Y}_{l,m}$, annihilate separately the two different sets of
doublets of the extended system. These mutually commuting integrals
generate a third, diagonal integral, ${\cal Z}_{l,m}$, that implies
the reflectionless property of the Hamiltonian: it appears as the
Lax integral, which together with the Hamiltonian forms the Lax
pair. The $ {\cal PT}$-operator emerges naturally as a valuable
symmetry for the integrals of motion in view of the fact that all of
them appear as $ {\cal PT}$-symmetric operators,  in the same way as
the Hamiltonian. Nevertheless, the odd order integral ${\cal
Y}_{l,m}$ reveals a finite number of pairs of complex conjugate
eigenvalues when acts on the bound states which correspond to
resonances with complex conjugate energy values in the Hermitian
potential counterpart. We can say therefore that for this integral
the $ {\cal PT}$-symmetry is spontaneously broken.

The ${\cal PT}$-symmetry is composed of the space inversion,
$\mathcal{P}$, and the time reversal, $\mathcal{T}$, operators,
which play a role of the intertwiners between the mutually displaced
components $H^+_{l,m}(x)$ and $H^-_{l,m}(x)=H^+_{l,m}(x+i\pi)$ of
the extended Hamiltonian. In addition to them and the half-period
displacement operators, there are other discrete intertwiners, the
products of which produce discrete symmetries of the extended
system. The commuting operators $\mathcal{R}_l:\,(l,m)\rightarrow
(-l,m)$ and $\mathcal{R}_m : (l,m) \rightarrow (l,-m-1)$ produce, particularly,  the
same effect as the differential intertwiners  $X^\pm_{l,m}$ and
$Y^\pm_{l,m}$, from which the supercharges $ {\cal X}_{l,m}$ and
${\cal Y}_{l,m}$ are composed. In addition to a usual choice for the
$\Z_2$-grading operator $\Gamma=\sigma_3$, other choices are also
possible. The product of the $\mathcal{P}$ and of the operator of
the displacement for the half of the imaginary period is one of
them, which also happens to be the grading operator for the hidden,
nonlinear bosonized supersymmetries of the subsystems $H^+_{l,m}$
and $H^-_{l,m}$, where the Lax operators $Z^+_{l,m}$ and $Z^-_{l,m}$
are identified as the odd supercharges.

On the other hand, the mentioned two sets of the discrete
eigenstates can be related between themselves by means of another
discrete symmetry of the Hamiltonian, which interchanges the
integer-valued lattice of the parameters $l$ and $m$ with the
half-integer-valued  lattice. It is only for such, integer or
half-integer, values of the parameters the complexified Scarf II
potentials are reflectionless. The indicated symmetry operation
intertwines the operators $A^{\pm}_{l,m}$ and $B^{\pm}_{l,m}$, which
are the building blocks for the intertwiners $X^\pm_{l,m}$ and
$Y^\pm_{l,m}$, respectively. As a consequence, the role of the
integrals ${\cal X}_{l,m}$ and ${\cal Y}_{l,m}$ is dually
interchanged by those specific discrete symmetries. In contrast with
the rest of the discrete symmetries, these duality generators have
no analog  in a form of differential operators.

The supersymmetric structure presented here displays several
similarities with the tri-supersymmetric structure in
self-isospectral Hermitian finite-gap systems with elliptic
potentials studied in \cite{trisusy}. In that class of Hermitian systems,
two distinct finite-dimensional representations of $sl(2,\R)$ are
realized on periodic and antiperiodic band-edge states; like here
three basic integrals of motion are present in the extended system,
and different choices for the grading operators are also possible.
The main difference with the present structure is that the
corresponding finite-gap elliptic systems are \emph{doubly
periodic}, in addition to the imaginary period the corresponding
systems  have also a \emph{real} period, and there the
self-isospectral systems are shifted for the half of their real
period. The corresponding tri-supersymmetric systems studied in
\cite{trisusy} are described by the associated Lam\'e potentials, which
constitute a subclass of the Darboux-Treibich-Verdier family
(\ref{dtv}). It is interesting therefore to investigate the question
of existence of tri-supersymmetric structure for such a class of
doubly periodic ${\cal PT}$-symmetric potentials, where the extended
Hamiltonian would unify the self-isospectral partners with a mutual
complex displacement.

In the definition of a physically consistent, positively definite
inner product for the systems with non-Hermitian Hamiltonians, the
existence of the operator ${\cal C}$ of the nature  of a charge
conjugation operator seems to be crucial \cite{BenderReview}. An open question
is the existence of such a  kind of the operator for the
complexified Scarf II potential. One can wonder then if the
supersymmetric structure discussed here can be helpful in this
sense, specifically, if the ${\cal C}$  can be expressed in terms
of, or related to the non-trivial integrals of motion.

Particular cases of the potential
 with $l\in\Z$ and $m=0$ (and equivalent cases
obtained by symmetry transformations of indexes) discussed here
appear in quantum field theory in curved space-times \cite{Maloney}.
A natural question is if a general case of the complex
reflectionless potential plays any
role in some related problems, and if the the revealed
supersymmetric structure could give some insight to these theories.

\vskip0.2cm

 \noindent \textbf{Acknowledgements.}
 The work has been partially
 supported by FONDECYT Grants 1095027 (MP)  and 3100123 (FC).
FC acknowledges also financial support via the CONICYT grants
79112034 and Anillo ACT-91: \textquotedblleft
Southern Theoretical Physics Laboratory\textquotedblright\ (STPLab). MP and FC are grateful, respectively, to CECs and
Universidad de Santiago de Chile for hospitality. The Centro de
Estudios Cient\'{\i}ficos (CECs) is funded by the Chilean Government
through the Centers of Excellence Base Financing Program of Conicyt.



\begin{thebibliography}{99}

\bibitem{B&B}
 C.~M.~Bender, S.~Boettcher,
 \emph{``Real spectra in non-Hermitian Hamiltonians having PT symmetry,''}
Phys.\ Rev.\ Lett.\  {\bf 80}, 5243 (1998), [arXiv:physics/9712001].

  \bibitem{BBM}
  C.~M.~Bender, S.~Boettcher and P.~Meisinger,
  \emph{``PT symmetric quantum mechanics,''}
  J.\ Math.\ Phys.\  {\bf 40}, 2201 (1999), [arXiv:quant-ph/9809072].

   \bibitem{DDT}
P.~Dorey, C.~Dunning and R.~Tateo, \emph{``Spectral equivalences,
Bethe Ansatz equations, and reality properties in PT-symmetric
quantum mechanics,''}
 J.\ Phys.\  A {\bf 34}, 5679 (2001), [arXiv:hep-th/0103051].


\bibitem{Mpseudoh}
 A.~Mostafazadeh,
  \emph{``Pseudo-Hermiticity versus PT symmetry.
  The necessary condition for the reality of the spectrum,''}
  J.\ Math.\ Phys.\  {\bf 43}, 205 (2002), [arXiv:math-ph/0107001];
  \emph{``Pseudo-Hermiticity versus PT symmetry 2. A Complete characterization of non-Hermitian Hamiltonians with a real spectrum,''}
  J.\ Math.\ Phys.\  {\bf 43}, 2814 (2002), [arXiv:math-ph/0110016].

 \bibitem{BenderReview}
  C.~M.~Bender,
  \emph{``Making sense of non-Hermitian Hamiltonians,''}
  Rept.\ Prog.\ Phys.\ \ {\bf 70}, 947  (2007), [hep-th/0703096 [hep-th]].

\bibitem{ReviewM}
  A.~Mostafazadeh,
  \emph{``Pseudo-Hermitian representation of quantum mechanics,''}
  Int.\ J.\ Geom.\ Meth.\ Mod.\ Phys.\  {\bf 7}, 1191 (2010), [arXiv:0810.5643 [quant-ph]].

\bibitem{witten}
  E.~Witten,
  \emph{``Dynamical Breaking Of Supersymmetry,''}
  Nucl.\ Phys.\ B {\bf 188}, 513 (1981).


\bibitem{susyrev1}
  F.~Cooper, A.~Khare and U.~Sukhatme,
  \emph{``Supersymmetry and quantum mechanics,''}
  Phys.\ Rept.\  {\bf 251}, 267 (1995), [arXiv:hep-th/9405029];
G.  Junker, \emph{Supersymmetric Methods in Quantum and Statistical
Physics}, (Springer, Berlin, 1996); Bagchi B.K.,
 \emph{Supersymmetry in quantum and classical mechanics},
 Chapman \& Hall/CRC Monographs and Surveys in Pure and
 Applied Mathematics, Vol. 116, (Chapman \& Hall/CRC, Boca Raton, FL, 2001).


\bibitem{MatSal}
V. B. Matveev and M. A. Salle, \emph{Darboux Transformations and
Solitons},  (Springer, Berlin, 1991).



\bibitem{nSUSY}
 A.~A.~Andrianov, M.~V.~Ioffe and V.~P.~Spiridonov,
\emph{``Higher derivative supersymmetry and the Witten index,''}
  Phys.\ Lett.\ A {\bf 174}, 273 (1993), [arXiv:hep-th/9303005];
  A.~A.~Andrianov, M.~V.~Ioffe and D.~N.~Nishnianidze,
 \emph{``Polynomial SUSY in quantum mechanics and second
derivative Darboux
 transformation,''}
 Phys.\ Lett.\  A {\bf 201}, 103 (1995), [arXiv:hep-th/9404120].


 \bibitem{MPhidnon}
   M.~S.~Plyushchay, \emph{``Hidden nonlinear supersymmetries
   in pure parabosonic systems,''}
  Int.\ J.\ Mod.\ Phys.\ A {\bf 15}, 3679 (2000), [arXiv:hep-th/9903130].

\bibitem{KlMP}
  S.~M.~Klishevich and M.~S.~Plyushchay,
\emph{``Nonlinear supersymmetry, quantum anomaly and quasi-exactly
solvable systems,''}
  Nucl.\ Phys.\  B {\bf 606}, 583 (2001), [arXiv:hep-th/0012023]


 \bibitem{bosonized}
   M.~S.~Plyushchay,
  \emph{``Deformed Heisenberg algebra, fractional spin fields and supersymmetry without fermions,''}
  Annals Phys.\  {\bf 245}, 339 (1996), [arXiv:hep-th/9601116];
  J.~Gamboa, M.~Plyushchay and J.~Zanelli,
  \emph{``Three aspects of bosonized supersymmetry and
  linear differential field equation with reflection,''}
  Nucl.\ Phys.\ B {\bf 543}, 447 (1999), [arXiv:hep-th/9808062].


  \bibitem{boso1}
  F.~Correa and M.~S.~Plyushchay,
  \emph{``Hidden supersymmetry in quantum bosonic systems,''}
  Annals Phys.\  {\bf 322}, 2493 (2007), [arXiv:hep-th/0605104].
    \bibitem{boso2}
   V.~Jakubsky, L.~-M.~Nieto and M.~S.~Plyushchay,
  \emph{``The origin of the hidden supersymmetry,''}
  Phys.\ Lett.\ B {\bf 692}, 51 (2010), [arXiv:1004.5489 [hep-th]].

  \bibitem{Dunne}
  G.~V.~Dunne and J.~Feinberg,
  \emph{``Self-isospectral periodic potentials and supersymmetric quantum mechanics,''}
  Phys.\ Rev.\ D {\bf 57}, 1271 (1998), [arXiv:hep-th/9706012].

  \bibitem{Fetal}
  D.~J.~Fernandez, B.~Mielnik, O.~Rosas-Ortiz and B.~F.~Samsonov,
  \emph{``The phenomenon of Darboux displacements,''}
  Phys.\ Lett.\ A {\bf 294}, 168 (2002), [arXiv:quant-ph/0302204].

\bibitem{FerNegNie}
 D.~J.~Fernandez, J.~Negro and L.~M.~Nieto,
``\emph{Second-order supersymmetric periodic potentials,}''
  Phys.\ Lett.\  A  {\bf 275}, 338 (2000).

  \bibitem{self}
B.F. Samsonov, M.L. Glasser, J. Negro and L.M. Nieto,
  ``\emph{Second-order Darboux displacements,}"
   J. Phys.  A {\bf 36}, 10053 (2003), [arXiv:quant-ph/0307146].

  \bibitem{trisusy}
  F.~Correa, V.~Jakubsky, L.~-M.~Nieto and M.~S.~Plyushchay,
  \emph{``Self-isospectrality, special supersymmetry, and their effect on the band structure,''}
  Phys.\ Rev.\ Lett.\  {\bf 101}, 030403 (2008), [arXiv:0801.1671 [hep-th]];
  F.~Correa, V.~Jakubsky and M.~S.~Plyushchay,
  \emph{``Finite-gap systems, tri-supersymmetry and self-isospectrality,''}
  J.\ Phys.\  A {\bf 41}, 485303 (2008), [arXiv:0806.1614 [hep-th]].

  \bibitem{trisusydelta}
    F.~Correa, L.~-M.~Nieto and M.~S.~Plyushchay,
  \emph{``Hidden nonlinear su($2|2$) superunitary symmetry
  of N=2 superextended 1D Dirac delta potential problem,''}
  Phys.\ Lett.\ B {\bf 659}, 746 (2008), [arXiv:0707.1393 [hep-th]].

    \bibitem{ABtrisusy}
   F.~Correa, H.~Falomir, V.~Jakubsky and M.~S.~Plyushchay,
  \emph{``Supersymmetries of the spin-1/2 particle in the field of magnetic vortex, and anyons,''}
  Annals Phys.\  {\bf 325}, 2653 (2010), [arXiv:1003.1434 [hep-th]].

   \bibitem{newself1}
  M.~S.~Plyushchay and L.~-M.~Nieto,
  \emph{``Self-isospectrality, mirror symmetry, and exotic nonlinear supersymmetry,''}
  Phys.\ Rev.\ D {\bf 82}, 065022 (2010), [arXiv:1007.1962 [hep-th]].
     \bibitem{newself2}
  M.~S.~Plyushchay, A.~Arancibia and L.~-M.~Nieto,
  \emph{``Exotic supersymmetry of the kink-antikink crystal, and the infinite period limit,''}
  Phys.\ Rev.\ D {\bf 83}, 065025 (2011), [arXiv:1012.4529 [hep-th]].
     \bibitem{newself3}
 A.~Arancibia and M.~S.~Plyushchay,
 \emph{``Extended supersymmetry of the self-isospectral crystalline
and soliton chains,''}, Phys. Rev. D {\bf 85}, 045018 (2012), arXiv:1111.0600 [hep-th].


 \bibitem{susypt1}
  F.~Cannata, G.~Junker and J.~Trost,
  \emph{``Schr\"odinger operators with complex potential but real spectrum,''}
  Phys.\ Lett.\ A {\bf 246}, 219 (1998), [arXiv:quant-ph/9805085].

  \bibitem{susypt2}
    A.~A.~Andrianov, F.~Cannata, J.~P.~Dedonder and M.~V.~Ioffe,
  \emph{``SUSY quantum mechanics with complex superpotentials and real energy spectra,''}
  Int.\ J.\ Mod.\ Phys.\ A {\bf 14}, 2675 (1999), [arXiv:quant-ph/9806019].

     \bibitem{susypt3}
    M.~Znojil,
  \emph{``Shape invariant potentials with PT symmetry,''}
  J.\ Phys.\  A {\bf 33}, L61 (2000), [arXiv:quant-ph/9911116].

     \bibitem{susypt4}
   B.~Bagchi, F.~Cannata and C.~Quesne,
  \emph{``PT symmetric sextic potentials,''}
  Phys.\ Lett.\ A {\bf 269}, 79 (2000), [arXiv:quant-ph/0003085].

\bibitem{Znojil2000}
  M.~Znojil, F.~Cannata, B.~Bagchi and R.~Roychoudhury,
  \emph{``Supersymmetry without hermiticity within PT symmetric quantum mechanics,''}
  Phys.\ Lett.\ B {\bf 483}, 284 (2000), [arXiv:hep-th/0003277].

     \bibitem{susypt5}
    B.~Bagchi, S.~Mallik and C.~Quesne,
  \emph{``Generating complex potentials with real eigenvalues in supersymmetric quantum mechanics,''}
  Int.\ J.\ Mod.\ Phys.\ A {\bf 16}, 2859 (2001), [arXiv:quant-ph/0102093].

\bibitem{Dorey}
  P.~Dorey, C.~Dunning and R.~Tateo,
  \emph{``Supersymmetry and the spontaneous breakdown of PT symmetry,''}
  J.\ Phys.\ A {\bf 34}, L391 (2001), [arXiv:hep-th/0104119].


  \bibitem{Mpseudosusy}
  A.~Mostafazadeh,
  \emph{``Pseudo-supersymmetric quantum mechanics and isospectral pseudoHermitian Hamiltonians,''}
  Nucl.\ Phys.\ B {\bf 640}, 419 (2002), [arXiv:math-ph/0203041].


 \bibitem{susynonlinearpt}
  S.~M.~Klishevich and M.~S.~Plyushchay,
 \emph{ ``Nonlinear holomorphic supersymmetry,
 Dolan-Grady relations and Onsager
  algebra,''}
  Nucl.\ Phys.\  B {\bf 628}, 217 (2002), [arXiv:hep-th/0112158];
   B.~Bagchi, S.~Mallik and C.~Quesne,
 \emph{  ``Complexified PSUSY and SSUSY interpretations of
  some PT symmetric Hamiltonians possessing two series of real energy
  eigenvalues,''}
  Int.\ J.\ Mod.\ Phys.\ A {\bf 17}, 51 (2002), [arXiv:quant-ph/0106021 [quant-ph]];
  A.~Khare and U.~Sukhatme,
  \emph{``Analytically solvable PT-invariant periodic potentials,''}
  Phys.\ Lett.\ A {\bf 324}, 406 (2004), [arXiv:quant-ph/0402106];
   A.~Sinha and P.~Roy,
  \emph{``PT symmetric models with nonlinear pseudosupersymmetry,''}
  J.\ Math.\ Phys.\  {\bf 46}, 032102 (2005), [arXiv:quant-ph/0505221];
 B.~F.~Samsonov,
  \emph{``Irreducible second order SUSY transformations between real and complex potentials,''}
  Phys.\ Lett.\ A {\bf 358}, 105 (2006), [arXiv:quant-ph/0602101];
  A.~Gonzalez-Lopez and T.~Tanaka,
  \emph{``Nonlinear pseudo-supersymmetry in the framework of N-fold supersymmetry,''}
  J.\ Phys.\ A {\bf 39}, 3715 (2006), [arXiv:quant-ph/0602177];
   A.~A.~Andrianov, F.~Cannata and A.~V.~Sokolov,
  \emph{``Non-linear supersymmetry for non-Hermitian,
  non-diagonalizable Hamiltonians. I. General properties,''}
  Nucl.\ Phys.\ B {\bf 773}, 107 (2007), [arXiv:math-ph/0610024].
  R.~Roychoudhury and B.~Roy,
  \emph{``Intertwining operator in nonlinear pseudo-supersymmetry,''}
  Phys.\ Lett.\ A {\bf 361}, 291 (2007);
  \emph{``Pseudo-supersymmetry and third order intertwining operator,''}
  Phys.\ Lett.\ A {\bf 372}, 997 (2008).


 \bibitem{GrNe}
  B. Bagchi and C. Quesne,
  \emph{``sl(2,$\mathbb{C}$) as a complex Lie algebra and
   the associated non-Hermitian
  Hamiltonians with real eigenvalues,''}
  Phys. Lett. A {\bf 273}, 285  (2000), [arXiv:math-ph/0008020].


\bibitem{BagQues}
  B.~Bagchi, C.~Quesne,
  \emph{``An update on PT-symmetric complexified Scarf II potential,
  spectral singularities and some remarks on
  the rationally-extended supersymmetric partners,''}
  J.\ Phys.\ A {\bf 43}, 305301 (2010), [arXiv:1002.4309 [quant-ph]].


\bibitem{Maloney}
 P.~Lagogiannis, A.~Maloney, Y.~Wang,
  \emph{``Odd-dimensional de Sitter space is transparent,''}
[arXiv:1106.2846 [hep-th]].


\bibitem{Wadati}
 M. Wadati,
\emph{``Construction of parity-time symmetric potential through the
soliton theory,''}
 Journ. of the Physical Society of Japan  {\bf 77},
074005 (2008).

\bibitem{Optics}
  Z.~H.~Musslimani, K.~G.~Makris, R.~El-Ganainy and D.~N.~Christodoulides,
  \emph{``Optical solitons in PT periodic potentials,''}
  Phys.\ Rev.\ Lett.\  {\bf 100}, 030402 (2008);
   \emph{``Analytical solutions to a class of
nonlinear Schr\"odinger equations with PT-like potentials,''} J.
Phys. A {\bf 41}, 244019 (2008); \emph{``PT-symmetric periodic
optical potentials,''}, Int. J. Theor. Phys.  {\bf 50},  1019
(2011).

\bibitem{Scarf2}
  J.~W.~Dabrowska, A.~Khare and U.~P.~Sukhatme,
  \emph{``Explicit wave functions for
  shape invariant potentials by operator techniques,''}
  J.\ Phys.\ A  {\bf 21}, L195 (1988).

\bibitem{LeZno}
  M.~Znojil and G.~Levai,
  \emph{``The interplay of supersymmetry and PT
  symmetry in quantum mechanics: a case study for the Scarf II potential,''}
  J.\ Phys.\  A {\bf 35}, 8793 (2002), [arXiv:quant-ph/0206013].

\bibitem{Ahmed}
  Z. Ahmed,
  \emph{``Real and complex discrete eigenvalues
  in an exactly solvable one-dimensional complex PT invariant potential,''}
  Phys.\ Lett.\ A {\bf 282}, 343 (2001).

\bibitem{LevaiCanVen2}
  G.~Levai, F.~Cannata and A.~Ventura,
  \emph{``Algebraic and scattering aspects of a PT symmetric solvable model,''}
  J.\ Phys.\  A {\bf 34}, 839 (2001).


 \bibitem{reflectionless}
 P. G. Drazin and R. S. Johnson, \emph{Solitons: An Introduction}, (Cambridge, UK:
Univ. Press, 1989);
A. Perelomov, Y. Zeldovich, \emph{Quantum Mechanics: Selected Topics},
World Scientific, Singapore, 1998.

 \bibitem{Coperator}
 C.~M.~Bender, D.~C.~Brody and H.~F.~Jones,
  \emph{``Complex extension of quantum mechanics,''}
  eConf C {\bf 0306234}, 617 (2003)
  [Phys.\ Rev.\ Lett.\  {\bf 89}, 270401 (2002)]
  [Erratum-ibid.\  {\bf 92}, 119902 (2004)],
  [arXiv:quant-ph/0208076].

\bibitem{KdV}
 F. Gesztesy and H. Holden, \emph{Soliton Equations and their
 Algebro-Geometric Solutions}, (Cambridge University Press, 2003).

 \bibitem{Lax}
  P.~D.~Lax,
  \emph{``Integrals of Nonlinear Equations of Evolution and Solitary Waves,''}
  Commun.\ Pure Appl.\ Math.\  {\bf 21}, 467 (1968).

\bibitem{Belo}
 E.~D.~Belokolos  et al, \emph{Algebro-geometric approach
tononlinear integrable equations}, (Springer, Berlin, 1994).

    \bibitem{PT}
    G.~P\"oschl and E.~Teller,
  \emph{``Bemerkungen zur Quantenmechanik des anharmonischen Oszillators,''}
  Z.\ Phys.\  {\bf 83}, 143 (1933).

  \bibitem{ptads}
  F.~Correa, V.~Jakubsky and M.~S.~Plyushchay,
  \emph{``Aharonov-Bohm effect on AdS(2) and nonlinear
  supersymmetry of reflectionless Poschl-Teller system,''}
  Annals Phys.\  {\bf 324}, 1078 (2009), [arXiv:0809.2854 [hep-th]].

  \bibitem{Veselov}
A. P. Veselov, \emph{``On Darboux-Treibich-Verdier potentials,''}
Lett. Math. Phys. {\bf 96}, 
209 (2011), [arXiv:1004.5355].

  \bibitem{asso}
W. Magnus and S. Winkler, \emph{HillÕs Equation} (Wiley, New York, 1966).

\bibitem{specialfunctions}
E. T. Whittaker and G. N. Watson, \emph{A course of modern
analysis}, (Cambridge Univ. Press, Cambridge, 1980), M. Abramowitz
and I. Stegun (Eds.), \emph{Handbook of Mathematical Functions},
Dover (1990).


\bibitem{AhmedAd}
  Z.~Ahmed,
  \emph{``Addendum to 'Real and complex discrete eigenvalues
   in an exactly solvable one-dimensional complex PT invariant potential',''}
  Phys.\ Lett.\ A {\bf 287}, 295 (2001).


\bibitem{bifurcationComment}
  B.~Bagchi, C.~Quesne,
\emph{``Comment on `Supersymmetry, PT-symmetry and spectral bifurcation',''}
  Annals Phys.\  {\bf 326}, 534 (2011),
  [arXiv:1007.3870 [math-ph]].
  K.~Abhinav and P.~K.~Panigrahi,
  \emph{``Supersymmetry, PT-symmetry and spectral bifurcation,''}
  Annals Phys.\  {\bf 325}, 1198 (2010), [arXiv:0910.2423 [quant-ph]].



\end{thebibliography}
\end{document}